\documentclass{article}
\usepackage{jfrExamplee}
\usepackage{graphicx}
\usepackage{natbib}
\usepackage{setspace}
\usepackage{algorithm}
\usepackage{algpseudocode}
\usepackage{amsmath}
\usepackage{amssymb}
\usepackage{xcolor}
\usepackage{subfig}

\usepackage{authblk}
\usepackage[bottom]{footmisc}

\newcommand{\Input}[1]{\State \textbf{Input:} #1}
\newcommand{\Output}[1]{\State \textbf{Output:} #1}
\DeclareMathOperator*{\argmin}{argmin}

\title{Precision yield estimation and mapping in manual strawberry harvesting with instrumented picking carts and a robust data processing pipeline}

\author{
\textbf{Uddhav Bhattarai}$^{1, *}$, \textbf{Rajkishan Arikapudi}$^1$ \textbf{Chen Peng}$^{1,\dagger}$, \textbf{Steven A. Fennimore}$^2$, \textbf{Frank N. Martin}$^3$, \textbf{Stavros G. Vougioukas}$^{1, *}$}

\small{
\affil{$^1$Department of Biological and Agricultural Engineering, University of California, Davis, CA, 95616, USA}
\affil{$^2$Department of Plant Sciences, University of California, Davis, CA, 95616, USA}
\affil{$^3$Crop Improvement and Protection Research Unit, U.S. Department of Agriculture Agricultural Research Service, Salinas, CA 93905, USA}
\affil{$^*$Corresponding authors: ubhattarai@ucdavis.edu, svougioukas@ucdavis.edu}
}

\begin{document}
\thispagestyle{plain}
\maketitle
\footnotetext{$\dagger$ Current affiliation: ZJU-Hangzhou Global Scientific and Technological Innovation Center, Zhejiang University, China}

\begin{abstract}
High-resolution yield maps for manually harvested crops are impractical to generate on commercial scales because yield monitors are available only for mechanical harvesters. However, precision crop management relies on accurately determining spatial and temporal yield variability. This study presents the development of an integrated system for precision yield estimation and mapping for manually harvested strawberries. Conventional strawberry picking carts were instrumented with a Global Positioning System (GPS) receiver, an Inertial Measurement Unit (IMU), and load cells to record real-time geo-tagged harvest data and cart motion. Extensive data were collected in two strawberry fields in California, USA, during a harvest season. To address the inconsistencies and errors caused by the sensors and the manual harvesting process, a robust data processing pipeline was developed by integrating supervised deep learning models with unsupervised algorithms. The pipeline was used to estimate the yield distribution and generate yield maps for season-long harvests at the desired grid resolution. The estimated yield distributions were used to calculate two metrics: the total mass harvested over specific row segments and the total mass of trays harvested. The metrics were compared to ground truth and achieved accuracies of 90.48\% and 94.05\%, respectively. Additionally, the accuracy of the estimated yield based on the number of trays harvested per cart for season-long harvest was better than 94\%. It showed a strong correlation (Pearson r = 0.99) with the actual number of counted trays in both fields. The proposed system provides a scalable and practical solution for specialty crops, assisting in efficient yield estimation and mapping, field management, and labor management for sustainable crop production.

\end{abstract}

\section{Introduction}
Yield estimation and mapping are critical in precision agriculture since they provide insights into spatial and temporal variability in yield that can be used to make informed decisions on field management, such as nutrient and fertilizer application, irrigation, and pest control. Accurate yield mapping requires precise yield data with high spatial resolution and temporal accuracy throughout harvesting. In highly mechanized crops, yield monitoring systems that measure and georeference the yield are commercially available and can be integrated into the harvesters. In contrast, fresh-market specialty crops such as apples, grapes, and strawberries are harvested manually. Hence, developing automated yield monitoring and mapping systems for these crops is challenging. The developed solutions must be compatible with the manual nature of harvesting the specific crop and must also be scalable, user-friendly, minimally disruptive to workers, adaptable, and robust.

For mechanized harvesting, yield estimation systems could be integrated into the harvesting machine \citep{maja2010development} or used as separate entities that work in conjunction with the harvesting machine \citep{JohnDeereActiveYield}. For example, \citet{maja2010development} developed an integrated citrus yield monitoring system for mechanical shake-catch harvesters. The system consisted of a GPS receiver for geolocation and a mass flow sensor with an impact plate and load cells to measure the impact force of citrus hitting the plate. In another work, \citet{jacques2018development} developed a vision-based yield monitoring system for a mechanical onion harvester.

Researchers have investigated several methods for estimating yields in manually harvested crops. These methods include pre-harvest data collection using vision-based technology, in-harvest data collection using instrumented tools such as specialized harvest bags or tracker devices, and post-harvest data collection by measuring the location and mass of collection bags or bins after harvesting. The pre-harvest indirect yield estimation methods are non-invasive and generally include a camera system to collect a sequence of ground-based or aerial imaging of the canopy, followed by algorithms to identify and track the object of interest and estimate yield \citep{nuske2014automated,roy2019vision}. For example, \citet{nuske2014automated} developed an invariant maximal detector followed by a region-growing algorithm to detect berries. Techniques such as image registration, camera position estimation, and voting schemes were used to merge detections and remove duplicate counts from multiple frames. Similarly, \citet{roy2019vision} used a Gaussian Mixture Model-based clustering approach to segment apples, followed by camera motion modeling to merge counts and remove duplicate counts. Additionally, \citet{chen2019strawberry} and \citet{di2019low} developed a vision-based pre-harvest yield mapping using aerial images. \citet{chen2019strawberry} developed a flower and fruit mapping system in strawberries using aerial images collected up to three meters above the ground level. The collected images were merged to create orthomosaic images fed to Faster R-CNN to detect and generate flower and fruit distribution maps. While the detection accuracy of the Faster R-CNN algorithm was promising, about 13.5\% of the flowers were not visible in the collected images due to occlusion. In another study by \citet{di2019low}, aerial vineyard images were captured to identify grape clusters and estimate the yield. Even if the survey was conducted in a partially defoliated vineyard, identifying green clusters was challenging due to occlusion and similarities in the appearance of leaves and clusters. While the advancements in machine vision and deep learning approaches have substantially improved, the accuracy of vision-based approaches is still severely affected by the outdoor unstructured environment, complete or partial fruit occlusions, variable lighting conditions, and ambiguous object appearance. 

In addition to pre-harvest yield estimation, some studies reported post-harvest yield estimation approaches in olives \citep{bayano2024olive} and oranges \citep{colacco2020yield}. \citet{bayano2024olive} developed a load cell/loading bolt system attached to a telescopic hydraulic arm to measure the mass of the bags of olives harvested from a mechanical trunk shaker and a Real-Time Kinematic Global Navigation Satellite System (RTK GNSS) system to georeference the bags. \citet{colacco2020yield} also georeferenced bags of harvested oranges; the mass was indirectly calculated. Researchers used a GPS to geolocate the bins or harvest bags and a tracker system, such as Radio Frequency Identification (RFID), for in-harvest real-time yield estimation. \citet{ampatzidis2009yield} combined RFID and Differential GPS (DGPS) to geolocate bins and associate them with pairs of peach trees. Similarly, \citet{bazzi2022yield} created a picker tracking device with an RFID reader and GPS for apple harvesting. Bag-emptying events into bins were identified through RFID scans. However, the approach could not offer precise yield maps because bag weight and exact location were not measured, and yield was estimated from bin and average bag weights. To compute accurate in-harvest yield maps, \citet{fei2017instrumented, fei2020estimation} developed instrumented apple picking bags, and \citet{anjom2018development} developed a prototype instrumented picking cart (aka \textit{iCarrito}) for strawberries with load cells, GPS, and IMU. Although the picking cart showed promise for precision yield estimation, the prototype was constructed using Polyvinyl Chloride (PVC) pipes, resulting in a bulky structure, the data were processed manually, and experiments were limited to small-scale harvest \citep{anjom2018development}. 

The main goal of this study was to build a yield monitoring system for manually harvested strawberries that is practical and low-cost so it can be widely adopted. Therefore, a key decision in system design was to use a low-cost GPS that relies on freely and widely available Satellite Based Augmentation System (SBAS) corrections. Although SBAS can typically achieve sub-meter accuracy, the spacing of the strawberry field furrows is small enough that occasional GPS errors make a cart - and the worker -  appear to be harvesting in a neighboring row instead of the actual harvest row. This neighboring row may not have been harvested yet, may have been harvested already, or may be harvested simultaneously by another worker. Typically, a worker harvesting in a row will not suddenly switch to a neighboring row or enter a row that is being harvested by someone else; however, there are instances where some pickers may do so. Hence, a key challenge is to develop a yield mapping algorithm that can recognize if the row corresponding to a worker's recorded location is correct or the result of GPS error, and correct it. 

Building upon previous work \citep{anjom2018development}, an improved version of an instrumented picking cart for strawberries was developed, and carts were deployed during commercial strawberry harvest in Santa Maria and Salinas, CA, over the 2024 harvest season. A yield estimation software pipeline was developed to process the harvest data, identify and correct inconsistencies in harvest location and row information, compute yield, and create a yield map for a typical harvest day and an entire season at a desired grid resolution. The following are the main contributions of the paper:
\begin{itemize}
    \item The typical picking patterns of workers during manual strawberry harvesting were modeled as mathematical inequalities that can be evaluated to detect and correct erroneous row assignments from noisy GPS data.
    \item A fully automated yield estimation pipeline was developed that processes and merges data from all available picking carts,  detects and corrects erroneous row assignments, and computes yields at desired grid resolution despite sensor errors and the inconsistent nature of manual harvesting.
    \item Extensive in-harvest data were collected during commercial strawberry harvest, and the dataset was made publicly available. Dataset link: [Public dataset link will be placed here]
\end{itemize}

The remaining sections of the paper are organized as follows. Section \ref{sec:preliminaries} discusses the background of manual strawberry harvesting and mathematically formulates the typical picking patterns followed by workers. Section \ref{sec:materialsmethods} discusses the development of the instrumented picking carts (iCarritos) and the comprehensive yield estimation pipeline. Section \ref{sec:resultsdiscussion} provides insights into the performance of the yield estimation and mapping system. Finally, Section \ref{sec:conclusion} concludes the paper by summarizing the key findings and contributions of the study. 

\section{Preliminaries}
\label{sec:preliminaries}
\subsection{Manual strawberry harvesting}
\begin{figure}[ht]
    \centering
    \includegraphics[width=0.3\textwidth]{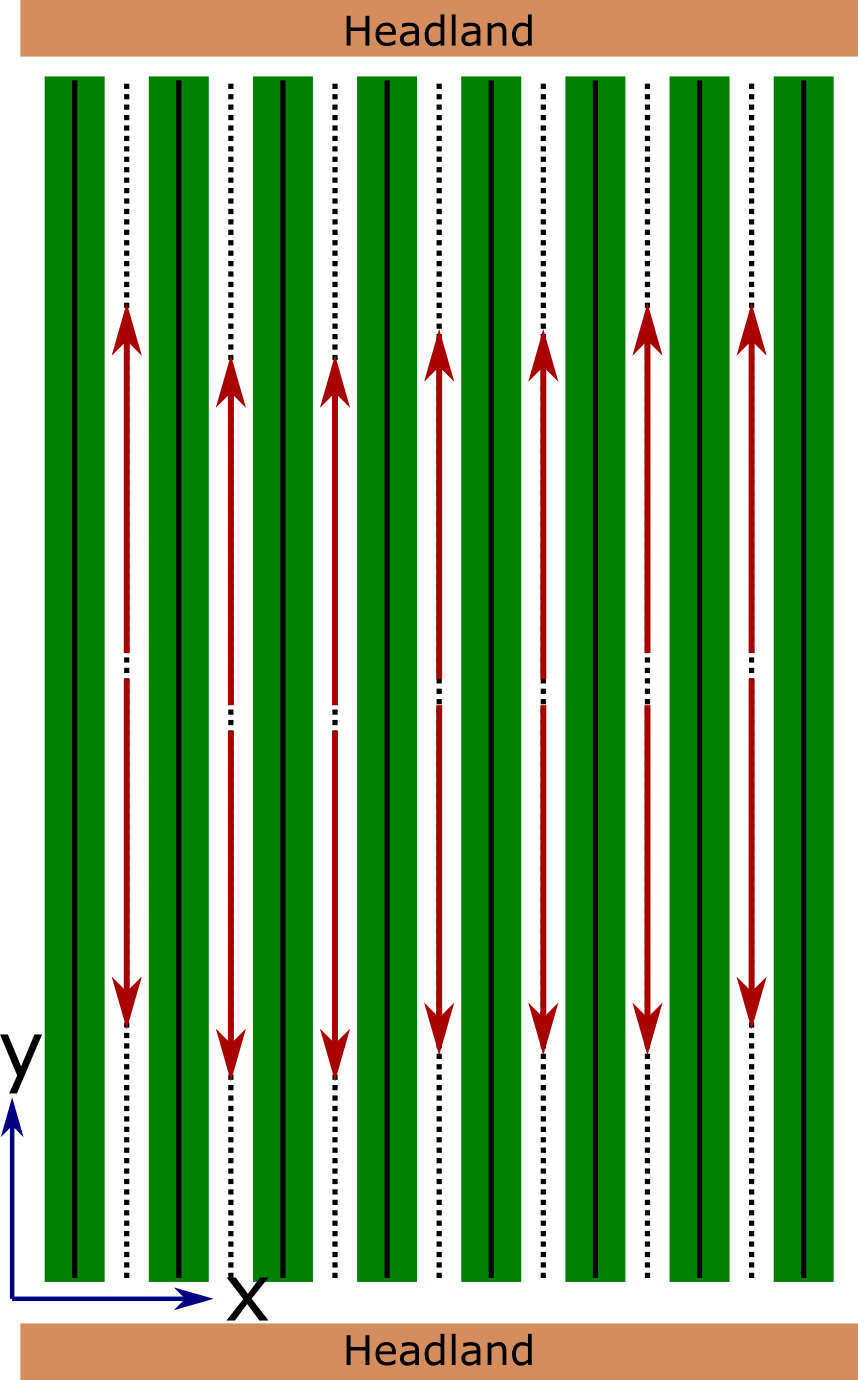}
    \caption{Layout of a typical strawberry field showing raised beds (green strips) and harvesting pattern. Red arrows indicate picker movement in the field during harvesting.}
    \label{fig:strawberry_field_layout}
\end{figure}

Strawberries are planted in parallel raised beds with rows/furrows between them for machine and worker traffic. Fruit collection and quality inspection stations are located at the headlands of the field. Headlands also facilitate transportation activities (see Figure \ref{fig:strawberry_field_layout}). The harvesting process begins with a group of picking crews, typically 15-40 pickers, depending on field size and crop yield. As the harvest starts, the picker crew starts from one side (e.g., the left side facing the field) and advances systematically towards the opposite end (e.g., the right side). Larger fields are often divided into two sections, typically upper and lower. Depending on field size, there may be two different crews, each harvesting the upper or lower section, or a single crew may sweep the entire field in a systematic pattern, for example, from the lower left to lower right and then upper right to the upper left. 

At the start of the harvest, each picker is assigned a carrito, four to eight clamshells, and a cardboard tray. Carritos are wireframe wheelbarrow picking carts (Figure \ref{fig:carritotoicarrito}, top-left) where pickers can place the cardboard tray holding clamshells with strawberries. The pickers usually begin harvesting from the approximate center of the row, progressing towards the headland (see Figure \ref{fig:strawberry_field_layout}). During harvesting, pickers pick berries from two beds on the left and right sides of the row. Furthermore, a picker will harvest an entire row before transitioning to another unharvested row. Additionally, a row is harvested by one picker at any given time, harvesting in one direction. If multiple pickers are in the same row, they must harvest in opposite directions. However, occasionally, a picker may "deviate" from the typical harvesting pattern.  

Furthermore, suppose the tray is full while the picker is inside a row. In that case, the picker walks to the edge of the field (headland) to deliver the full tray to the inspection and collection station and picks up an empty tray and clamshells to continue harvesting in the same row. However, if a tray is partially full when a picker reaches the end of the row, the picker moves to a new unharvested row and continues harvesting. This process continues cyclically until the end of the harvest, and the pickers are generally paid based on the number of trays they harvest (piece rate). \citet{seyyedhasani2020collaboration} modeled manual strawberry harvesting using a finite state machine formalism. The activities of picker $p$ during manual harvesting were grouped into discrete
operating states $s_p^i$ \\
\begin{equation}
\label{eq:pickerstates}
s_p^i \in \left\{
\begin{aligned}
&\text{start}, \text{idle-in-queue}, \text{walk-empty-tray-headland}, \text{walk-empty-tray-row}, \text{picking}, \text{walk-to-next-row}, \\
&\text{setup}, \text{transp-full-tray-row}, \text{transp-full-tray-headland}, \text{stop}
\end{aligned}
\right\}
\end{equation}

The picker states include picking and non-picking activities, such as switching rows, delivering trays, walking, and waiting. Although pickers are in different states, they follow a typical picking pattern for systematic harvesting. The following section mathematically presents the typical picking pattern followed by pickers during commercial strawberry harvest in detail.

\subsection{Mathematical formulation of typical picking pattern in manual strawberry harvesting}
Let $P = \{p_i\}_{i=1}^P$ be the set of pickers, each assigned with instrumented picking carts $C$ for harvesting strawberries in a field with rows $R = {(r_{x_j}, r_{y_j})}_{j=1}^M$. Let \(\sigma(p, t_1, t_2, r, d)\) a binary variable representing the harvesting activity of picker \(p\) in row \(r\) in a direction \(d\) over the time interval \([t_1, t_2]\). The picker travel direction \(d \in \{1, -1\}\) denotes picker movement within a row during the harvesting with \(1\) indicating forward movement and \(-1\) indicating backward movement. The binary variable $\sigma$ models the spatiotemporal signature of the pickers defined as:

\begin{equation}
    \sigma(p, t_1, t_2, r, d) =
    \begin{cases} 
        1, & \text{if picker } p \text{ is harvesting in row } r \text{ in direction } d \text{ between times } t_1 \text{ and } t_2, \\
        0, & \text{otherwise.}
    \end{cases}
\end{equation}

The binary variable $\sigma$ was used to model the typical picking behaviors followed by workers during strawberry harvest as algebraic inequalities. Our goal was to evaluate the inequalities using the carrito data and thus detect and correct errors in the data because of the sensor error or atypical picker behavior.

\begin{enumerate}

    \item \textbf{Row completion}: While harvesting a specific row, pickers complete the entire row before transitioning to another row. Intermittent switching between rows is not a general practice to ensure efficiency and consistency in harvesting. A picker acts this way if the following condition is met:
    
    \begin{equation}
        \sum_{r \in R} \sum_{d \in \{1, -1\}} \sigma(p, t_1, t_2, r, d) \leq 1, \quad \forall p \in P, \forall t_1 < t_2
    \end{equation}

    \item \textbf{Row occupancy}: A row can only be occupied by one picker at any given time, harvesting in one direction. If multiple pickers are in the same row, they harvest in opposite directions. This will be true if the following condition is met:
    \begin{equation}
        \sum_{p \in P} \sigma(p, t_1, t_2, r, d) \leq 1, \quad \forall r \in R, \forall d \in \{1, -1\}, \forall t_1 < t_2
    \end{equation}

\end{enumerate}

\section{Materials and Methods}
\label{sec:materialsmethods}
\subsection{Instrumented Picking Carts}
\begin{figure}
    \centering
    \includegraphics[width=0.75\linewidth]{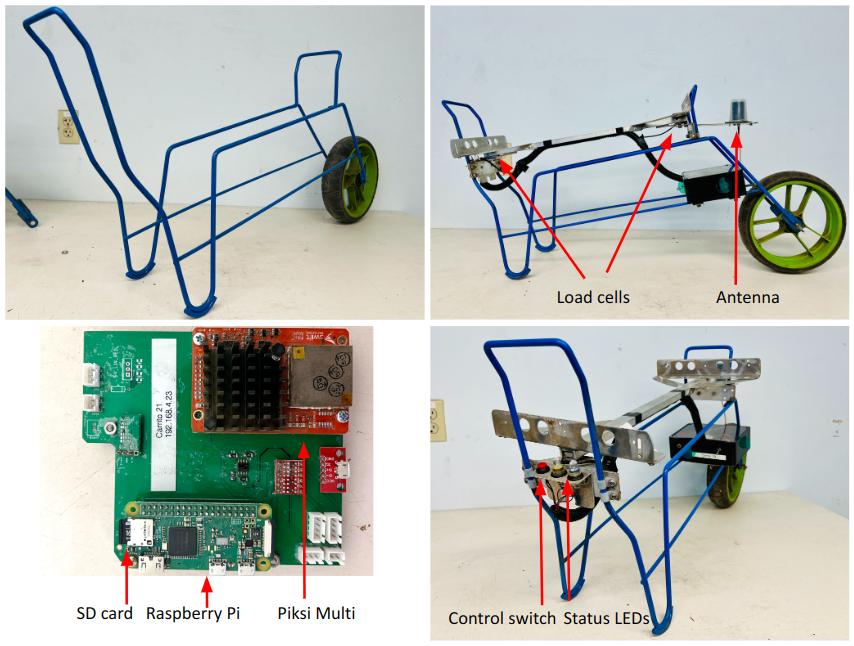}
    \caption{Evolution of standard carrito to iCarrito. (Top-Left) The wireframe structure of a typical commercial carrito. (Top-Right) Instrumented carrito showing GNSS antenna, load cells, and control box. (Bottom-Left) Close-up view of the main circuit board featuring Raspberry Pi 0W, SwiftNav Piksi Multi GNSS unit, and SD card. (Bottom-Right) Instrumented carrito control switch, status LEDs}
    \label{fig:carritotoicarrito}
\end{figure}

In this study, conventional picking carts (aka \textit{carritos}) were instrumented with sensors, mounting systems, and control hardware to develop the instrumented carritos (also referred to as "iCarritos"). Figure \ref{fig:carritotoicarrito} shows the modification of a carrito into iCarrito. Two load cells were installed at the front and back of the carrito for precise harvest mass measurement, Figure \ref{fig:carritotoicarrito} (Top-Right). Furthermore, a Piksi Multi (Swift Navigation Inc., USA) GNSS unit with integrated IMU  was installed to measure location (latitude, longitude) and acceleration data along x, y, and z axes ($a_x$, $a_y$, $a_z$). A Raspberry Pi 0W (Raspberry Pi Foundation, UK) microcomputer was used as the central processing unit, and an SD card was used to run the carrito software and store the data during the harvest. The control system consisted of one main power toggle switch connected directly to the battery, one control switch for selecting operation modes, and two status LEDs for system feedback. 

The iCarrito was programmed to operate in two primary modes: data collection mode and data upload mode. The data collection mode was used to gather real-time mass, location, and motion data during harvesting. On the other hand, the data upload mode was used to transfer the collected data to a local computer via Wi-Fi for data management and yield analysis. The data was collected at the rate of 10Hz with information on Raspberry Pi Unix timestamp, GNSS Unix timestamp, GNSS time of week ($ms$), latitude ($degrees$), longitude ($degrees$), height ($m$), acceleration in x-axis ($m/s^2$), acceleration in y-axis ($m/s^2$), acceleration in z-axis ($m/s^2$), raw mass ($Kg$). At the start of the harvest, data were only collected after the GNSS unit obtained SBAS corrections (horizontal Circular Error Probable accuracy was 0.75 meters). Furthermore, no data was collected when the SBAS signal was lost during harvesting.

\begin{figure}
    \centering
    \includegraphics[width=0.1\linewidth]{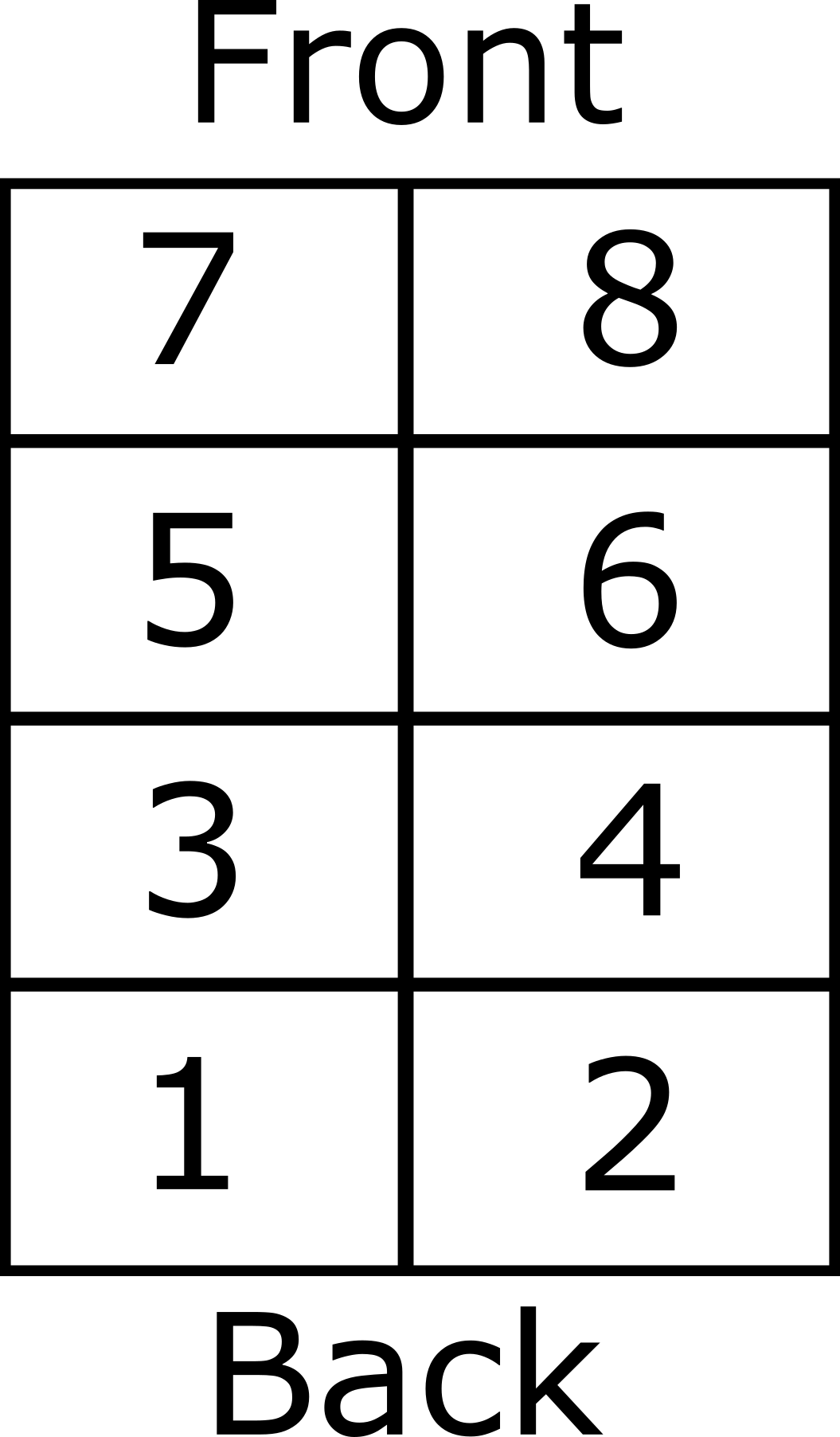}
    \caption{Sequence of filling the cardboard tray with the clamshells filled with predetermined mass for load cell calibration.}
    \label{fig:carrito_calibration}
\end{figure}

\textbf{Load cell calibration}: To accurately correspond the load cell signals to the harvested mass, each cart's load cells were calibrated before the carts were taken to the field for the first time. The load cells were calibrated using the method described by \citet{anjom2018development}. Objects with known weight were packed in clamshells and placed in a tray on a cart in a sequence as shown in Figure \ref{fig:carrito_calibration}. Once the data points were gathered, a simple linear regression was performed to correlate the measured mass with the load cell's voltage readings. The resulting calibration parameters (slope and intercept) were saved to a file for future mass calculations during harvesting. The harvest data was checked for errors after each harvest throughout the season, and the load cells were re-calibrated if necessary.

\subsection{Experimental Study Site}


\begin{figure}[ht]
    \centering
    \includegraphics[width=0.43\textwidth]{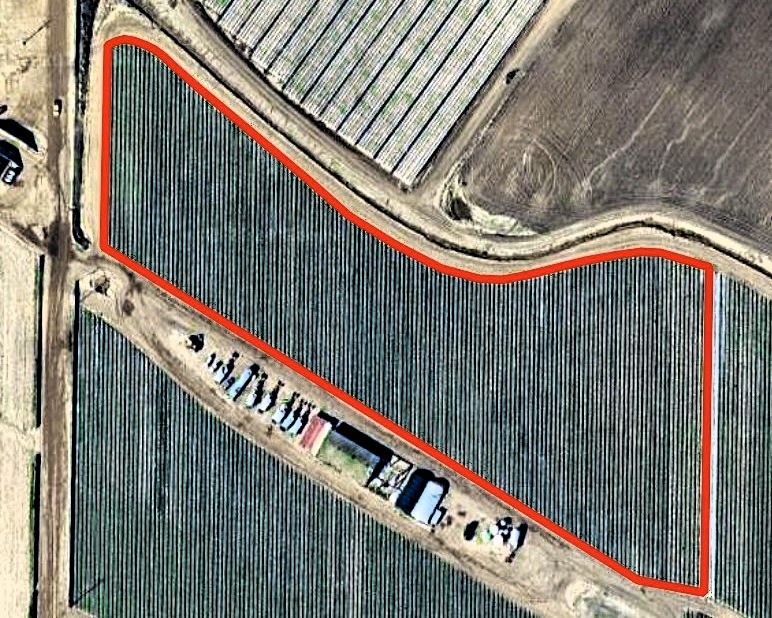}
    \includegraphics[width=0.43\textwidth]{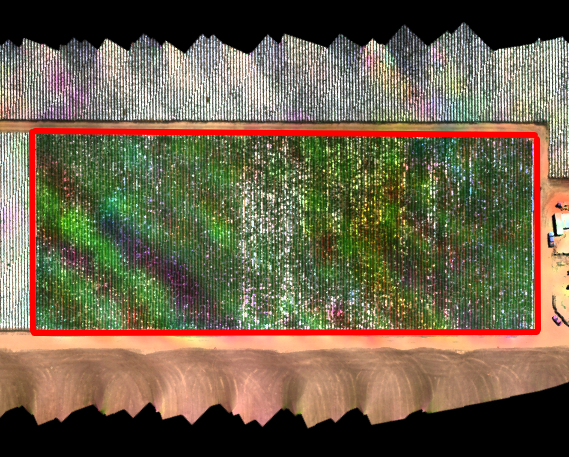}
    \includegraphics[width=0.43\textwidth]{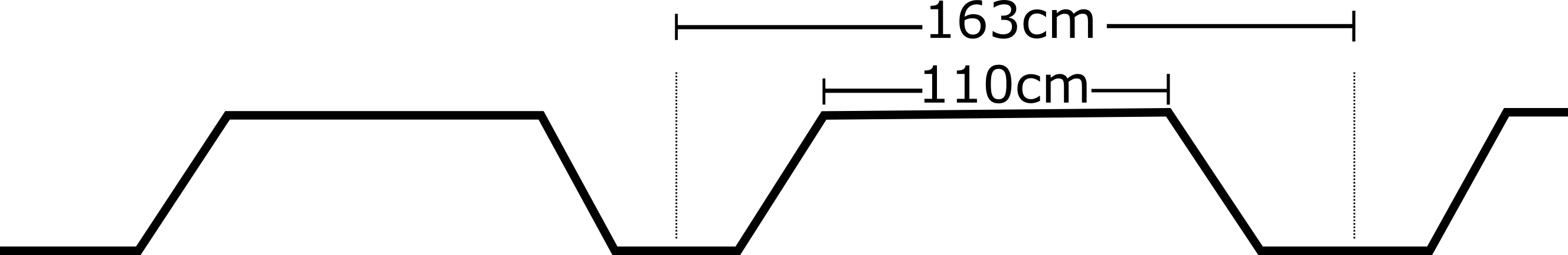}
    \includegraphics[width=0.43\textwidth]{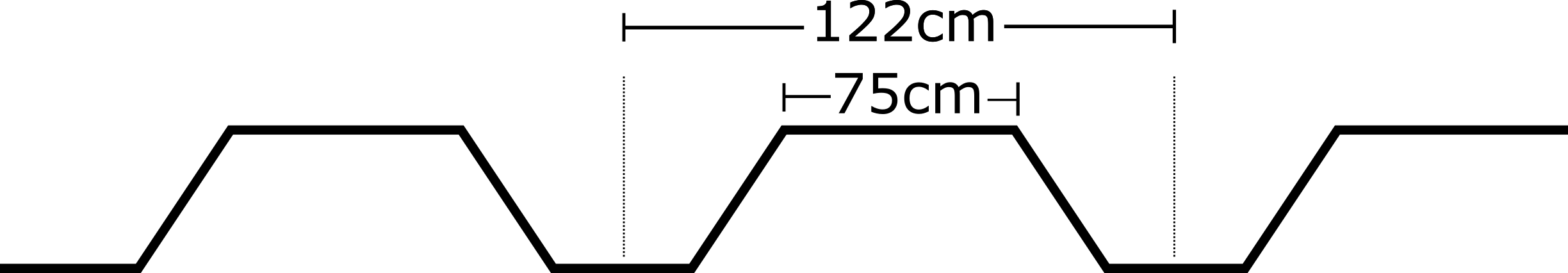}
    \caption{Aerial view and row specifications of study sites in California in 2024 season. (Left) Santa Maria farm (Area: $17,862.15 m^2$) growing Fronteras variety. (Right) Salinas farm (Area: $16,560.56 m^2$) growing Cabrillo variety. Red polygons identify harvest regions}
    \label{fig:experimental_site}
\end{figure}

\begin{figure}[ht]
    \centering
    \includegraphics[width=0.5\textwidth]{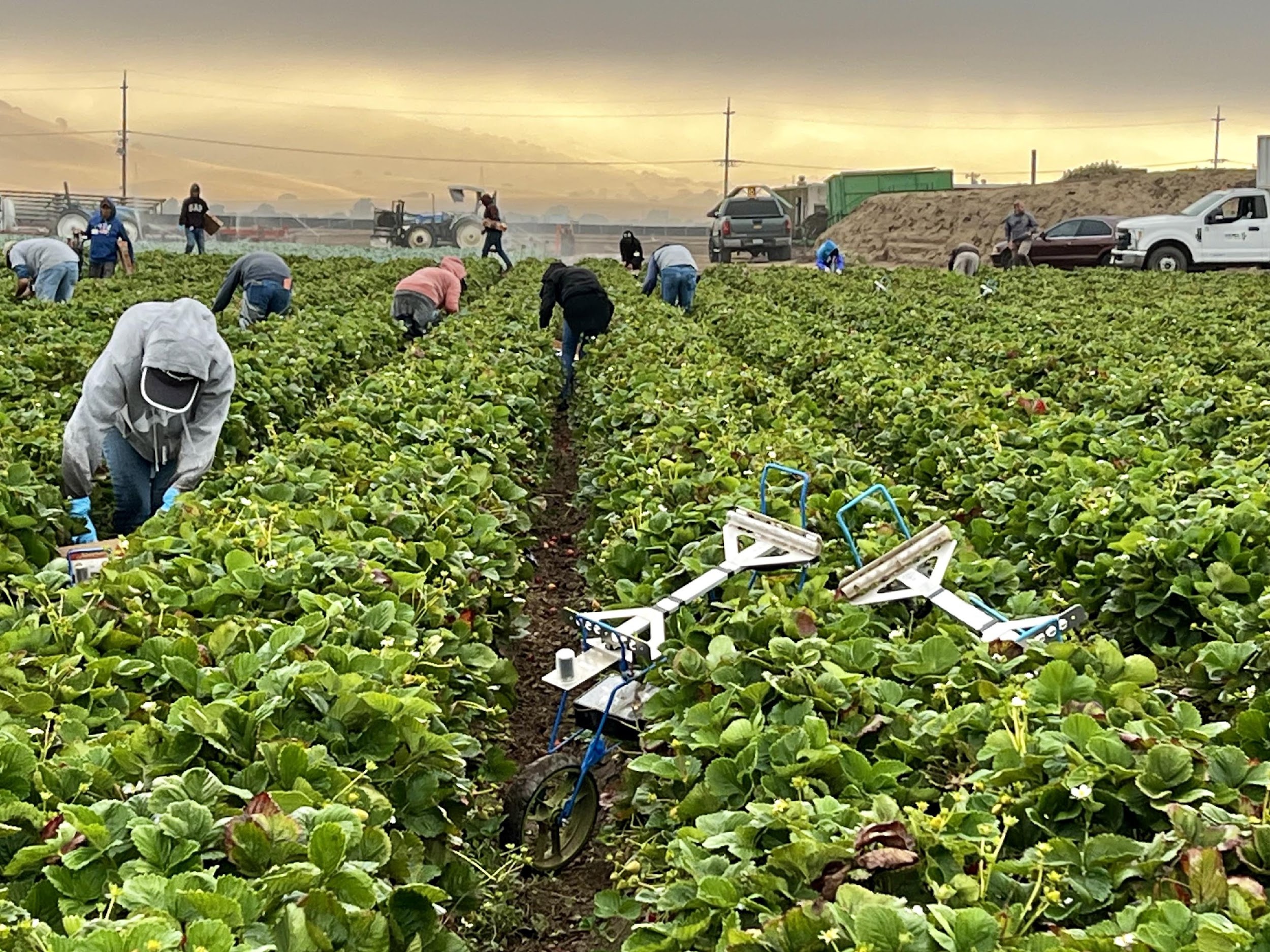}
    \caption{Field deployment of the iCarritos during commercial strawberry harvest in Salinas, California. Two separate crews were deployed to harvest the upper and lower sections of the field as identified by the direction they were facing}
    \label{fig:carrito_field_deployment}
\end{figure}

Studies were conducted in two major strawberry-producing regions in California during the 2024 strawberry growing season, as shown in Figure \ref{fig:experimental_site}. The selected fields were a commercial strawberry field (Central West Produce) in Santa Maria (Location: $34^\circ 53' 59.8"" \text{N}, 120^\circ 28' 25.0"" \text{W}$, Area: $17,862.15 m^2$) growing Fronteras strawberry, and a commercial strawberry farm (Gema Berry Farms) in Salinas, CA (Location: $36^\circ 37' 35.1"" \text{N}, 121^\circ 32' 16.4"" \text{W}$, Area: $16,560.56 m^2$), growing Cabrillo variety.  The raised bed width in Santa Maria was larger, with a bed width of 110 cm, compared to Salinas, with a bed width of 75 cm. The raised bed centers were also pre-mapped using the RTK-GPS to compute row locations. The row locations were calculated as the center of the neighboring bed centers, resulting in row spacings of 163 cm in Santa Maria and 122 cm in Salinas. Figure \ref{fig:carrito_field_deployment} shows a typical harvest day with pickers harvesting strawberries in Salinas field with two separate crews harvesting the upper and lower sections of the field.

\begin{figure}[ht]
    \centering
    \includegraphics[width=0.45\linewidth]{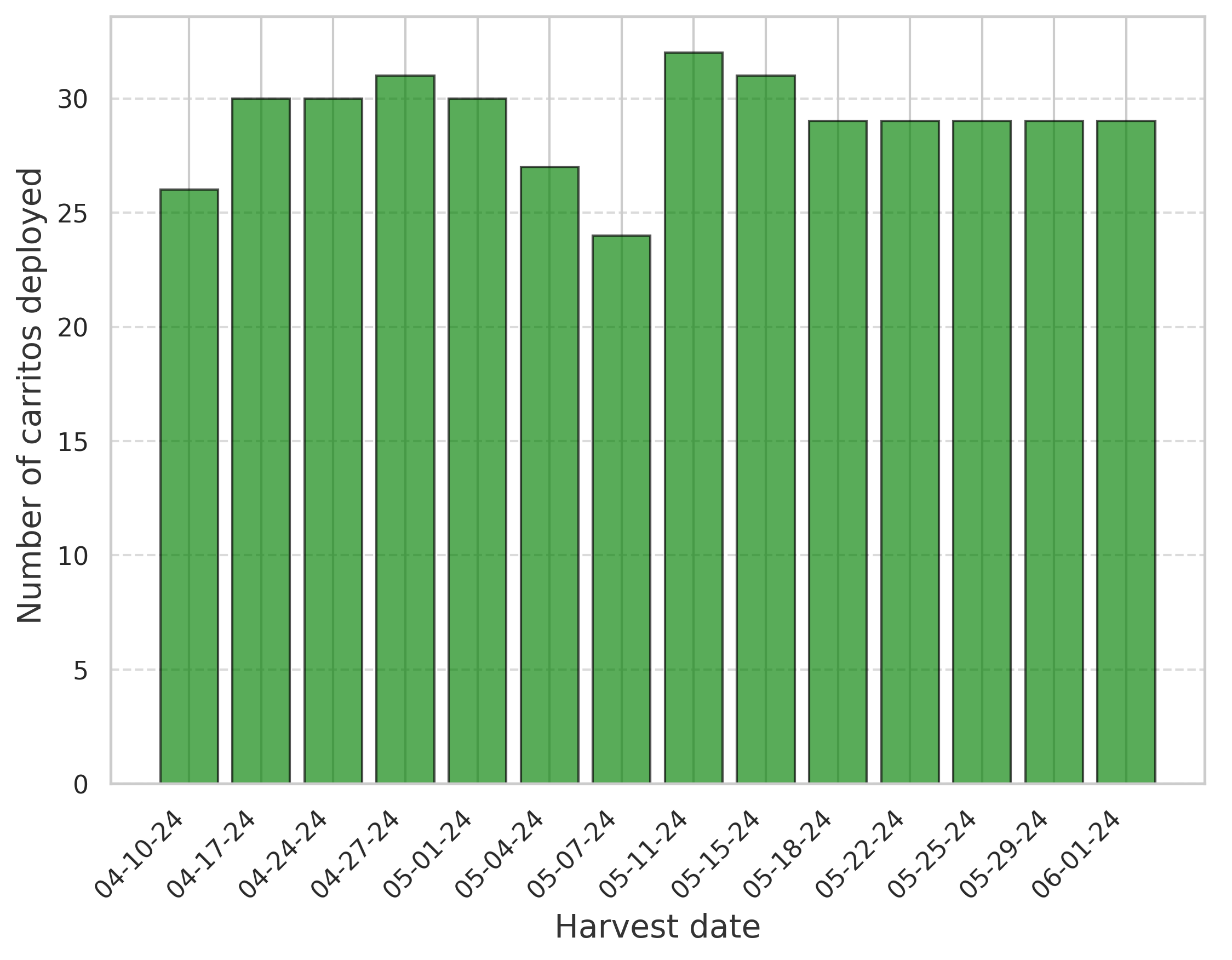}
    \includegraphics[width=0.45\linewidth]{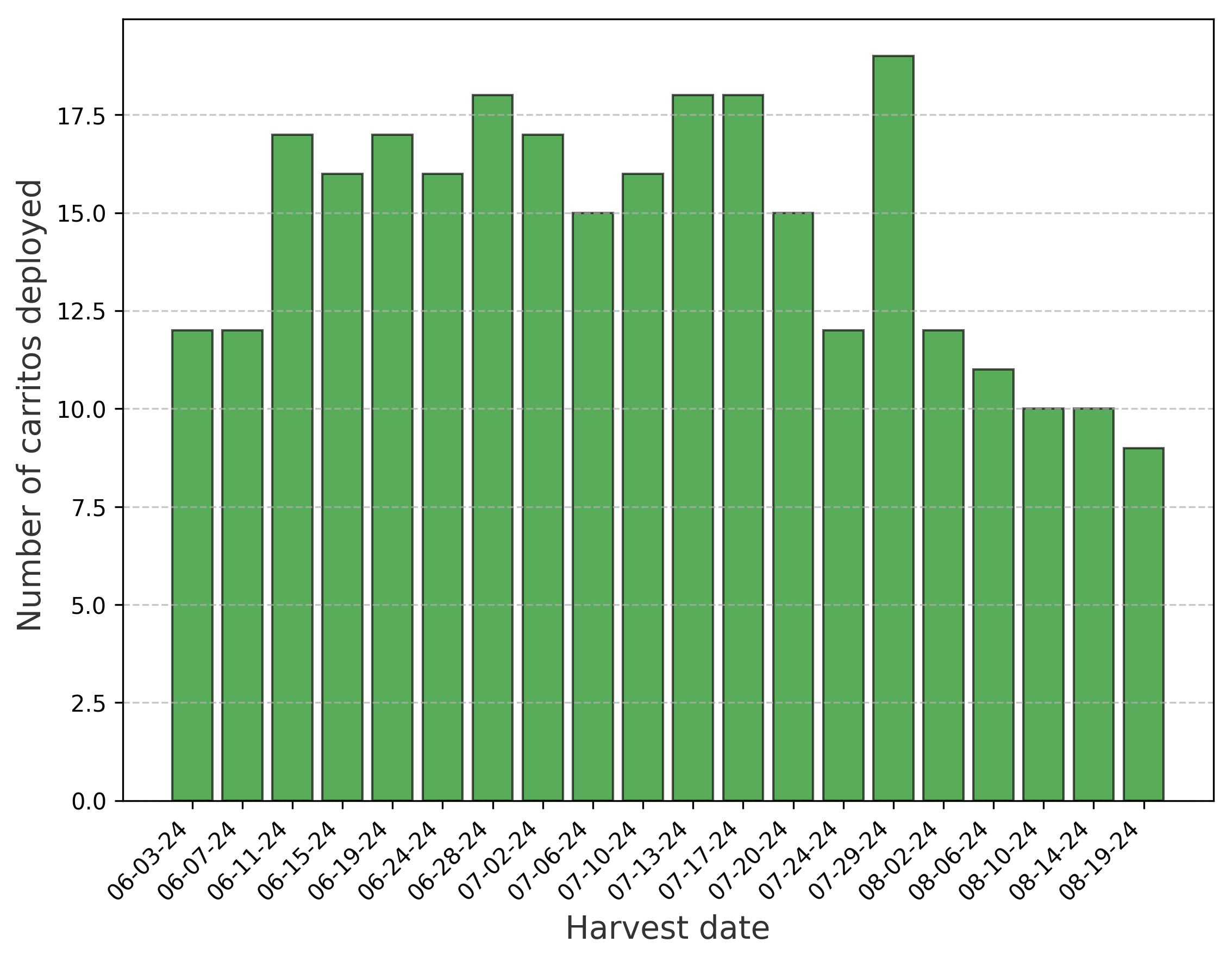}
    \caption{(Left) The number of iCarritos deployed each harvest day across 14 harvest sessions in Santa Maria with a mean deployment of 29 carritos per day. (Right) The number of iCarritos deployed across all (20) harvest sessions in Salinas with a mean deployment of 15 carritos per day.}
    \label{fig:santamaria_salinas_field_deployed}
\end{figure}

Figure \ref{fig:santamaria_salinas_field_deployed} shows the distribution of the iCarritos deployed in both field locations. Successive harvests were conducted every three to seven days, and the grower determined the appropriate interval based on yield and maturity. Although the Santa Maria field was harvested 27 times through the entire season, only 14 harvest session data were processed for yield analysis in this work because of the large volume of data. Meanwhile, the Salinas field was harvested 20 times, with an average of 15 carritos deployed daily. Due to disease and low yield, the grower was forced to stop harvesting in Salinas in the middle of the harvesting season. In Santa Maria, each carrito harvested an average of 35 trays daily, ranging from 3 to 106 trays, resulting in a total of 14282 trays. In Salinas, each carrito harvested an average of 29 trays per day, ranging from 3 to 90 trays, with a total of 11254 harvested trays.

The mathematical formulation was used to model the expected picker movements during systematic harvesting. However, some of the data recorded during the carritos' field deployment did not align with the anticipated data due to the limitation of the sensing system and/or the handling of the carts by the pickers. As discussed earlier, our goal was to develop a low-cost scalable system that used an SBAS-based localization system without Real-Time Kinematic (RTK) correction. The integrated Piksi multi GNSS unit has rated horizontal positional accuracy of approximately 75 cm \citep{piksispecs}. Although this level of GPS accuracy was sufficient given the row spacings of 163 cm in Santa Maria and 122 cm in Salinas, the system still experienced location errors, mainly when the satellite connection was weak. Moreover, substantial variations and abrupt motions were observed in the way pickers handled the carts. Since pickers were compensated based on the number of trays harvested, there was a tendency to rush movements during the harvest. Different strategies could be employed to address the deviation from the standard picking behavior violations, such as using a more accurate and robust RTK-GPS system and training the pickers to standardize the handling of the picking carts. While RTK-GPS can reduce positional errors, it may lead to bulkier, costlier carts, and training is challenging due to the seasonal and transient nature of the workforce. Hence, we prioritized developing robust algorithms to process noisy data from low-cost, practical sensing and data collection systems.

\subsection{Overview of the Yield Estimation Pipeline}

\begin{figure}
    \centering
    \includegraphics[width=0.98\linewidth]{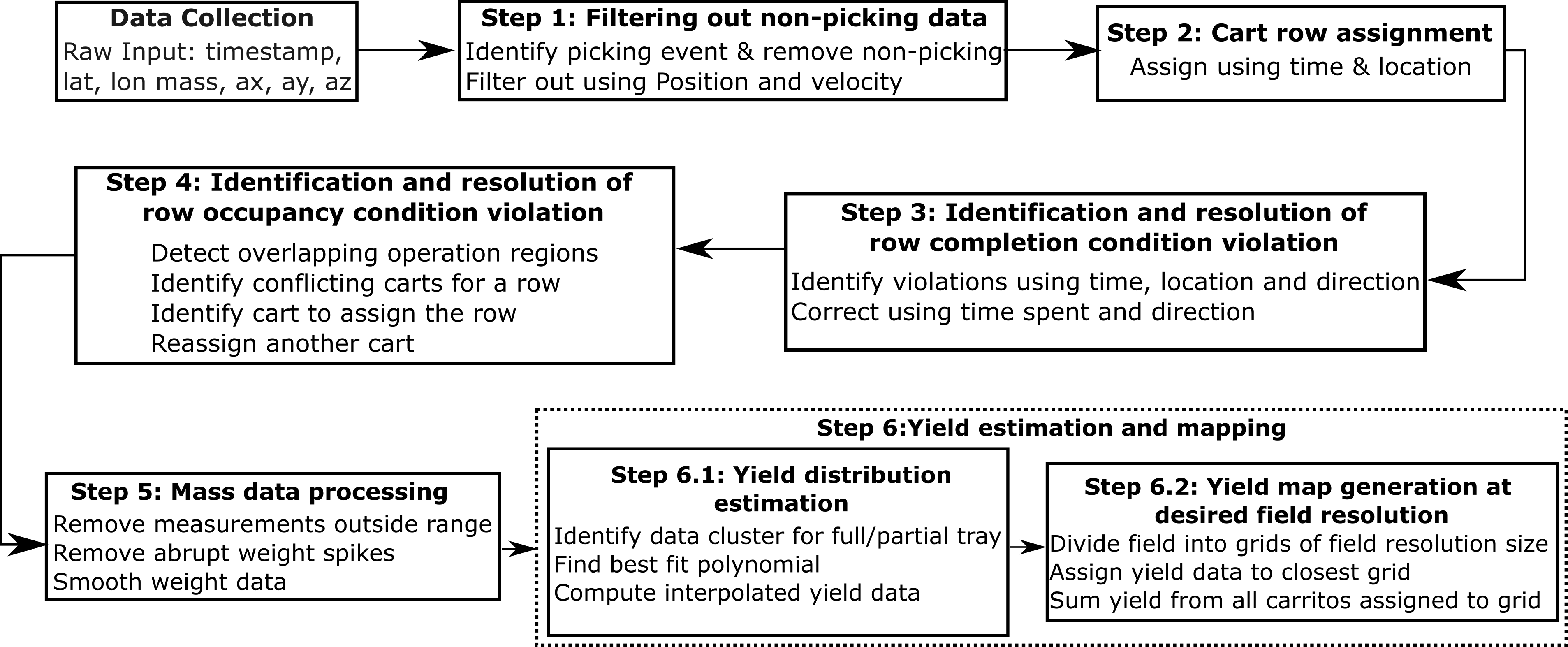}
    \caption{Flow diagram of the yield estimation pipeline consisting of six processing steps. Processing steps include data collection, filtering out non-picking data, cart row assignment, identification and correction of row completion and row occupancy condition violation, harvest mass processing, and yield estimation and mapping.}
    \label{fig:carrito_data_processing}
\end{figure}

Figure \ref{fig:carrito_data_processing} illustrates the overall flow of the yield estimation pipeline starting from raw harvest data collected from the iCarritos. The raw data was noisy due to sensor limitations and the variability in the pickers' handling of the carts. A CNN-LSTM-based model was developed and deployed to filter out the irrelevant non-picking data by classifying the raw data into two categories: data collected during the picking or a non-picking activity. Furthermore, cart locations that fell outside the field boundary were removed. After filtering out the irrelevant data, the cart locations were assigned to one of the pre-mapped rows by clustering the harvest data based on time and location and assigning them to the closest row based on Euclidean distance. A transformation matrix was computed based on the bed center location from RTK GPS to transform iCarrito GPS coordinates to local field coordinates to simplify the location data processing. When necessary, the field coordinates could be converted back into GPS coordinates through an inverse transformation.

The following steps involved identifying and correcting row completion and occupancy condition violations. Timestamps, location data, and harvest direction were used to identify violations of the row completion condition. If a violation was detected with a cart operating in multiple rows at a time, the cart locations were corrected such that a cart was harvesting a single row at a time. Furthermore, to identify the violation of row occupancy conditions, an algorithm was developed to identify overlapping regions in rows where multiple carts were observed harvesting simultaneously. If the violation was detected, the conflicting row was assigned to the appropriate cart, while the other cart was reassigned to one of the neighboring rows.

After addressing the location condition violation, the mass data was processed. First, the mass measurements that fell outside the normal harvest mass range were removed. Next, abrupt mass spikes that occurred due to the handling of the trays or berries, as well as any unintended pressure applied to the cart, were filtered out. Furthermore, the median filter was used to remove noise and reduce the impact of outliers. The final step was divided into two phases:  yield distribution estimation and yield map generation. 

In the first phase, yield distribution was estimated using data interpolation so that the mass data could be sampled uniformly. During harvest, pickers harvest all the berries within reach and then push the cart forward for some distance before resuming harvest. This movement results in a staircase-like mass signal, making uniform mass sampling challenging. To address this, harvest data related to each partial and full tray were identified, and a polynomial model was fitted to the location and mass data to find the best-fit polynomial. This polynomial was then used to calculate the interpolated yield data at uniform intervals of one foot. 

In the second -  yield mapping - phase, the field was divided into grid cells at a desired resolution, and the interpolated yield data from each carrito were assigned to the nearest grid cell. Finally, the total yield at each grid cell was computed by combining yield data from all the carritos for that harvest day whose data were assigned to each grid cell.

\subsubsection{Step 1: Filtering out non-picking data}
\begin{figure}[ht]
    \centering
    \includegraphics[width=0.48\linewidth]{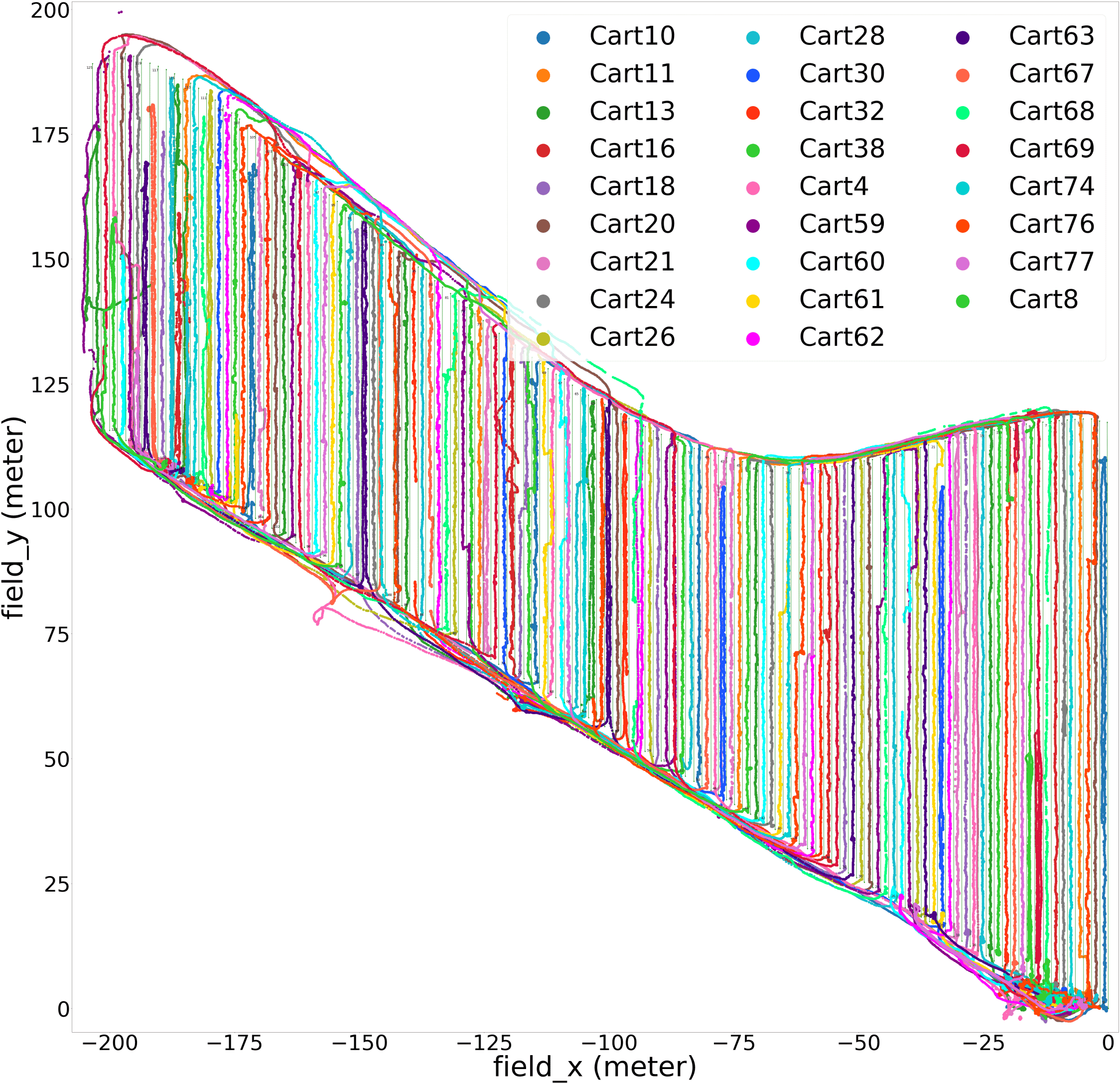}
    \includegraphics[width=0.5\linewidth]{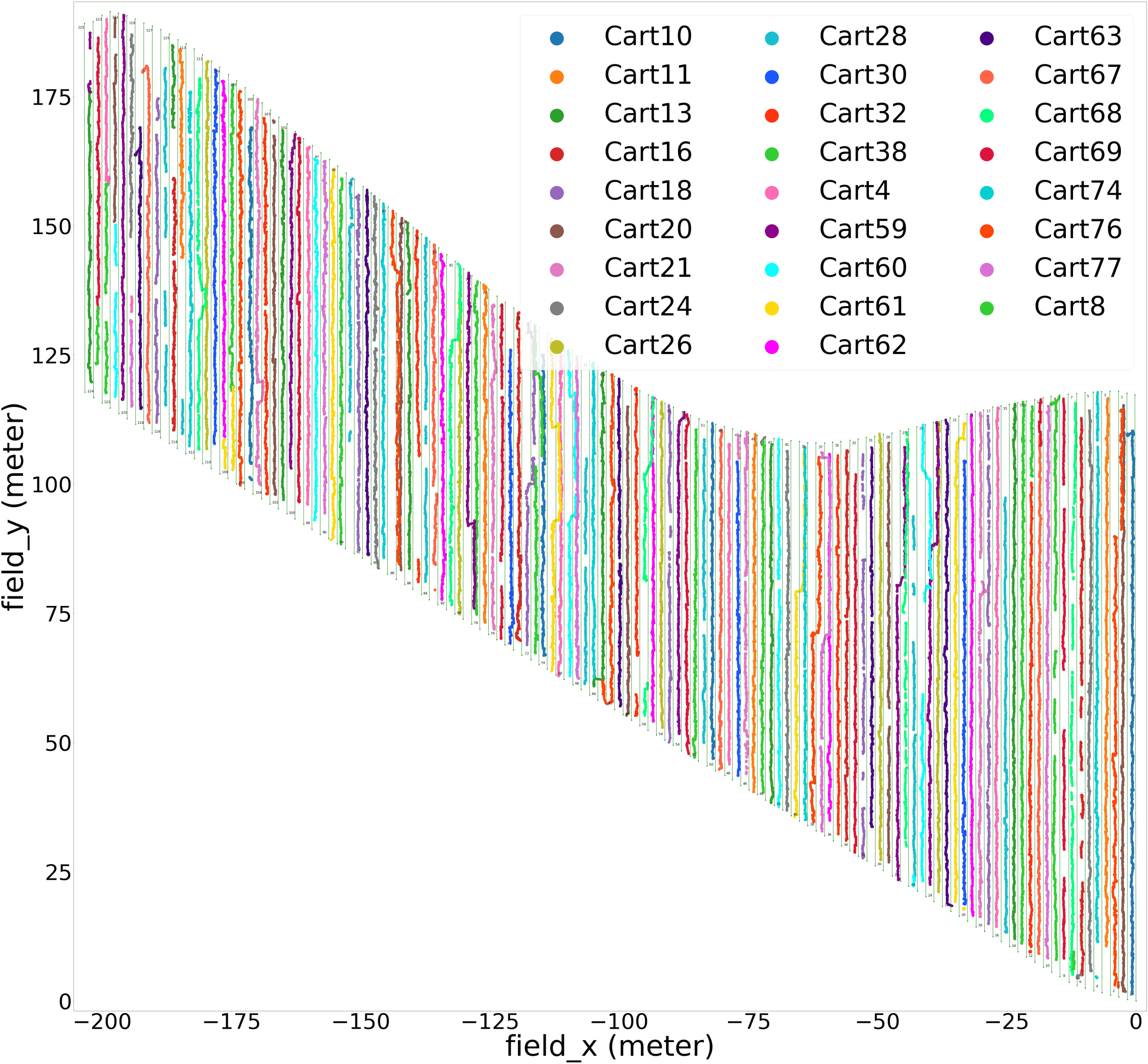}
    \caption{Cart location traces before and after Step 1: Filtering non-picking data. (Left) Raw data shows data related to non-picking activities and abrupt location jumps. Colored traces represent unique cart identifiers. (Right) Filtered traces after applying Step 1 showing clean harvest data.}
    \label{fig:before_after_filter}
\end{figure}

This step aimed to identify and keep the picking data for yield estimation and mapping by filtering out irrelevant non-picking data from the raw harvest data. Because of the GPS error and manual nature of the harvesting, the raw data often contained noise and abrupt location jumps, as shown in Figure \ref{fig:before_after_filter} (Left). Furthermore, while the different picker operation states (outlined in equation \ref{eq:pickerstates}) could be helpful in applications such as robot-assisted crop transportation systems, only valid picking data was necessary to estimate the yield. Additionally, any recorded data outside of the picking area was also irrelevant. A robust deep learning model was developed to remove noise and irrelevant data by combining a Convolution Neural Network (CNN) and Long Short-term Term Memory (LSTM). Any remaining data points that may have been missed by CNN-LSTM and recorded outside of the picking area were also removed. Figure \ref{fig:before_after_filter} (Right) shows the clean cart traces after removing irrelevant data.

Since the CNN-LSTM is a fully supervised deep learning network, raw harvest data was annotated into the "Pick" and "NoPick" classes. The annotated data was then divided into windows containing 100 data points (10 seconds of harvest activity) and used as ground truth for training the network. The network was trained for up to 50 epochs using a categorical cross-entropy loss cost function. Experiments were conducted to systematically evaluate the impact of input features mass, acceleration, and velocity on model performance. It was found that the network achieved high accuracy and robustness with an F1 score of 0.973 when trained using mass and acceleration data. Hence, it was selected to filter out irrelevant data in this study. Furthermore, mass was the most prominent feature in classifying harvest data, mainly due to the increasing mass signal during picking. A detailed description of the algorithm's architecture and experiments are presented in \citet{bhattarai2025data}.

\subsubsection{Step 2: Cart row assignment}
\label{sec:cart row assignment}
\begin{figure}
    \centering
    \includegraphics[width=0.48\linewidth]{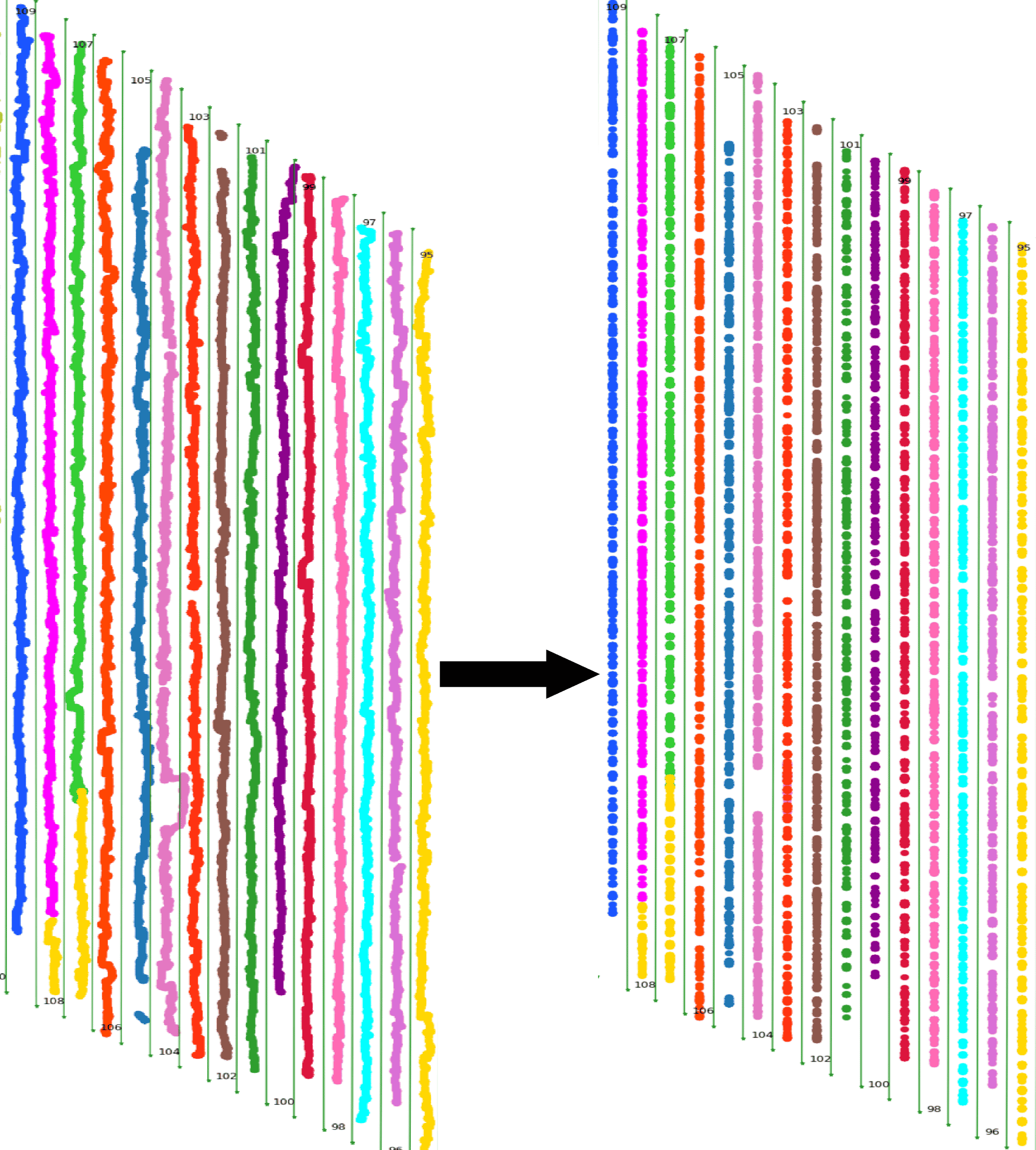}
    \caption{Cart location traces before and after Step 2: Cart row assignment. (Left) Cart traces showing pickers' imperfect (non-straight) harvest paths. (Right) Cart traces form straight lines after being assigned to the pre-mapped row. }
    \label{fig: before and after cart row assignment}
\end{figure}

The objective of this step was to assign each recorded carrito location to a pre-mapped row in the field. As shown in Figure \ref{fig: before and after cart row assignment} (Left), the carrito locations did not form perfectly straight lines, as they were pushed inside the rows. The reason is the manual nature of the work and the noise in the GPS signals. Mapping the carrito traces to pre-mapped rows standardizes their locations to the row center and eliminates irrelevant natural movement variations without impacting yield analysis. Algorithm \ref{alg:row_assignment} illustrates the overall process of row assignment. Let the trajectory of cart $(C)$ be $\{(x_i, y_i, t_i)\}_{i=1}^N$ where $(x_i, y_i)$ are cart locations in the local field coordinate with an associated timestamp of $t_i$. Let the row locations were $R = {(r_{x_j}, r_{y_j})}_{j=1}^M$. Where $M$ was the total number of rows. The goal was to develop a mapping function $f: C \rightarrow \{1, 2, ..., M, \emptyset\}$ that assigned each filtered data point to a row or classify as noise ($\emptyset$). The cart locations were first grouped using Density-Based Spatial Clustering of Applications with Noise (DBSCAN) clustering. The main idea was that if the data points were close enough in time and space, they could be considered part of the same movement or trajectory. The grouped data points were then assigned to a particular row based on the proximity of the median point of the cluster to a row. Figure \ref{fig: before and after cart row assignment} (Right) shows the assigned cart locations to each row, where the wavy cart path was translated into straight lines.

\begin{algorithm}
\caption{Row assignment of cart locations}\label{alg:row_assignment}
\footnotesize
\begin{algorithmic}[1]
\Input{$C = \{(x_i, y_i, t_i)\}_{i=1}^N$, $R = {(r_{x_j}, r_{y_j})}_{j=1}^M$}
\Output{$C'$ (Updated $C$ with corrected row assignments)}

\State $clusters \gets \text{DBSCAN}(C, \epsilon, minPts)$
\For{$cluser_{id}$ in $clusters$ \textbf{where} $cluster_{id} \neq noise_{id}$}
        \State $x_{row} \gets \arg\min_{r_x \in R}(|median([x | (x,y,t) \in cluster_{id}]) - r_x|)$
        \For{$point$ in $cluster_{id}$} 
            \State $C' \gets C.append(x_{row})$
        \EndFor
\EndFor

\State \Return $C'$ 

\end{algorithmic}
\end{algorithm}

\subsubsection{Step 3: Identification and resolution of row completion condition violation}
\label{sec:row_completion_correction}
\begin{figure}
    \centering
    \includegraphics[width=0.25\linewidth]{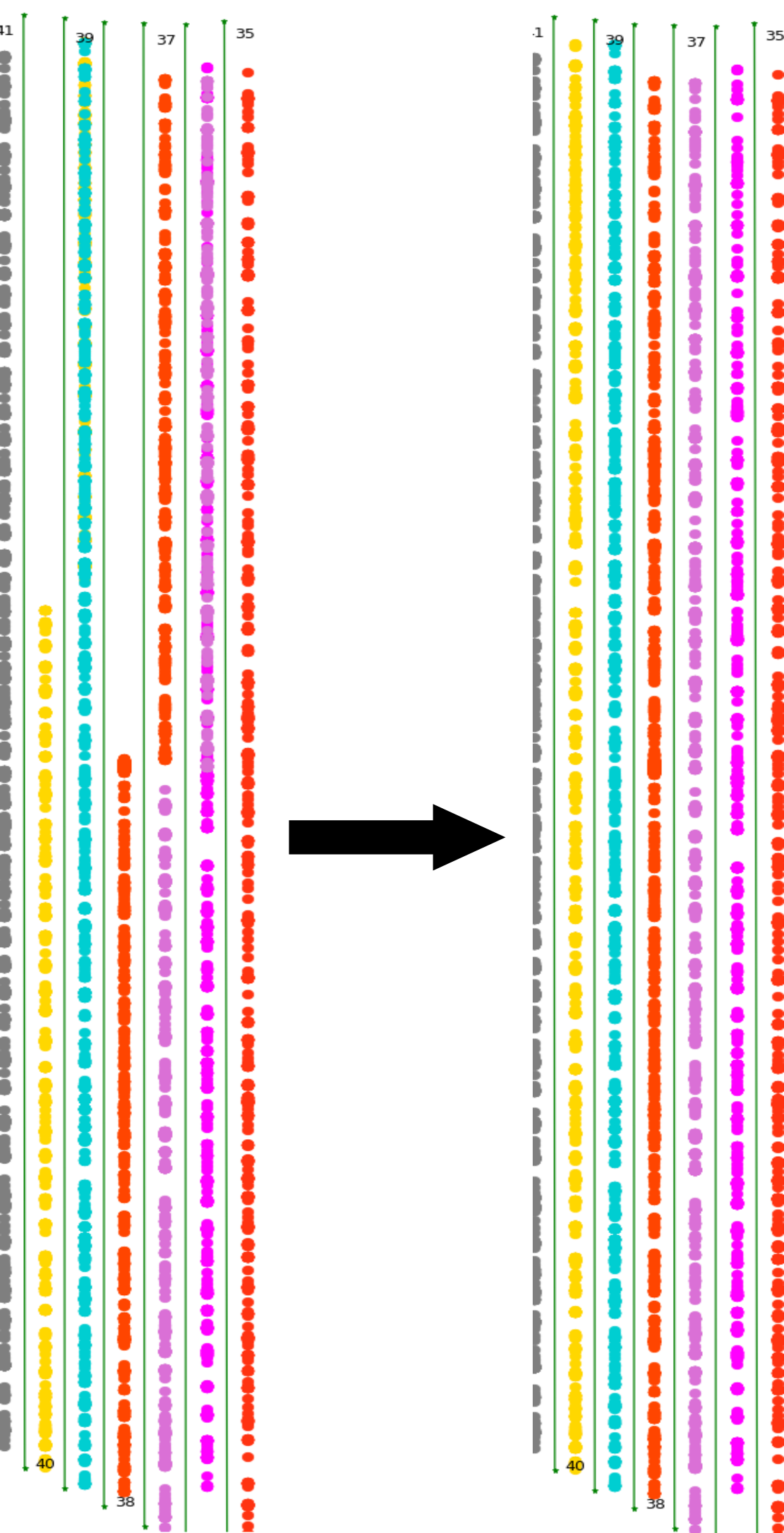}
    \caption{Mapped cart location traces before and after Step 3: Identification and resolution of row completion condition violation. (Left) Cart traces are incorrectly mapped to multiple rows due to GPS errors and picker movement. About half of the yellow and red colored cart traces were incorrectly assigned to neighboring rows on the right. (Right) Corrected cart traces after the row completion violation correction, showing proper assignment to single rows.}
    \label{fig:beforeafterrowcompletioncorrection}
\end{figure}

This step aimed to identify and resolve violations of the row completion condition by re-mapping carts incorrectly assigned to multiple rows to a single row. Assignment of cart data to multiple neighboring rows could occur if the picker was harvesting in two different sections of the field while traveling in opposite directions (which does not violate the condition) or if the picker was traveling in the same direction and the location was incorrectly recorded in multiple rows, violating the condition. As observed in Section \ref{sec:cart row assignment}, pickers did not move in a straight line. This movement pattern, combined with the GPS error and the manual nature of harvesting, may cause the cart location to appear as being in a neighboring row, resulting in cart traces being incorrectly mapped to multiple rows, as shown in yellow and red colored traces in Figure \ref{fig:beforeafterrowcompletioncorrection} (Left). This step utilizes information on cart location, time spent harvesting in a row, and travel direction to identify and resolve the violations. Algorithm \ref{alg:row_correction} illustrates the process of determining the violation of the row completion condition and the steps taken to correct erroneous multiple-row assignments. For the given cart data $C = \{(x_i, y_i, t_i, x_{row})\}_{i=1}^N$, the assigned row information $\{ x_{row})\}_{i=1}^N$ was used to determine if a cart was assigned to pairs or triplets of rows. If a cart was assigned to pairs or triplets of the rows, the cart travel direction in each row was computed based on cart location data ($Y = \{(y_i)\}_{i=1}^N$) in $y$ direction. 

\begin{algorithm}
\caption{Identification and resolution of row completion condition violation}
\label{alg:row_correction}
\footnotesize
\begin{algorithmic}[1]
\Input{$C = \{(x_i, y_i, t_i, x_{\text{row}_i})\}_{i=1}^N$, $R = \{(x_{r_j}, y_{r_j})\}_{j=1}^M$}
\Output{$C'$ (Updated $C$ with corrected row assignments)}

\Function{AdjustMinorDataRow}{$C, r_{\text{min}}, r_{\text{maj}}$}
    \State $C' \gets C$
    \For{$i \gets 1$ to $|C'|$}
        \State $C'[i].x_{\text{row\_new}} \gets \begin{cases}
            r_{\text{maj}} & \text{if } C'[i].x_{\text{row}} = r_{\text{min}} \\
            C'[i].x_{\text{row}} & \text{otherwise}
        \end{cases}$
        \State $C'[i].\text{flag} \gets (C'[i].x_{\text{row}} = r_{\text{min}})$
    \EndFor
    \State \Return $C'$
\EndFunction

\Function{ProcessRowGroup}{$\text{rows}$}
    \State $\text{directions} \gets [\,]$
    \For{$r \in \text{rows}$}
        \State $Y_r \gets \{y_i \mid (x_i, y_i, t_i, x_{\text{row}_i}) \in C \wedge x_{\text{row}_i} = r\}$
        \State $\text{directions}.\text{append}(\text{ComputeTravelDirection}(Y_r))$
    \EndFor
    
    \For{$\text{AllEqual}(\text{directions})$}
        \State $r_{\text{maj}} \gets \text{GetMajorRow}(\text{rows}, C)$
        \For{$r \in \text{rows} \setminus \{r_{\text{maj}}\}$}
            \State $C' \gets \text{AdjustMinorDataRow}(C, r, r_{\text{maj}})$
        \EndFor
    \EndFor
\EndFunction

\State $\text{row}_{\text{set}} \gets \text{SortedUnique}(\{x_{\text{row}_i}\}_{i=1}^N)$
\State $R_{\text{pairs}}, R_{\text{triplets}} \gets \text{FindConsecutiveRows}(\text{row}_{\text{set}})$

\For{$\text{pair} \in R_{\text{pairs}}$}
    \State $C' \gets$ \Call{ProcessRowGroup}{$\text{pair}$}
\EndFor

\For{$\text{triplet} \in R_{\text{triplets}}$}
    \State  $C' \gets$  \Call{ProcessRowGroup}{$\text{triplet}$}
\EndFor

\State \Return $C'$

\end{algorithmic}
\end{algorithm}

\begin{figure}
    \centering
    \includegraphics[width=0.9\linewidth]{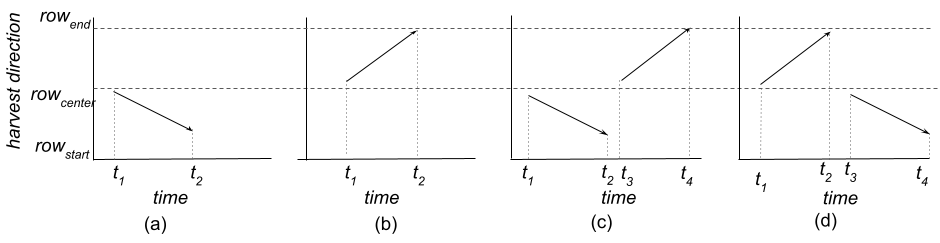}
    \caption{Representation of picker movement patterns during harvesting. The dashed horizontal line indicates the field center reference point. Arrows indicate the direction of picker movement. (a) Harvest direction from the field center to the start of the row (t1 to t2). (b) Harvest direction from the field center to the end of row (t1 to t2). (c),(d) Bidirectional picker movement where the same row was harvested from opposite directions at different time intervals (t1 to t2 and t3 to t4). }
    \label{fig:travel_dir}
\end{figure}
Various picker movement scenarios were considered to compute the picker travel direction. As illustrated in Figure \ref{fig:travel_dir}, the pickers begin harvesting from the center of the field and move towards the headland. Ideally, the travel direction could determined by determining the relative movement from the initial registered cart location to the final registered location over an interval of t1 to t2. However, in some cases where the crew size was small, pickers harvested the same row twice, once from one side of the field and then again from the opposite side. The time interval between these actions, t2 to t3 in Figure \ref{fig:travel_dir}, was stochastic and depended on the pickers' speeds and the size of the field. Furthermore, if the row lengths are small, a picker may harvest an entire row from the start to the end. Considering different picker movement scenarios, Algorithm \ref{alg:travel_dir} was developed, which used peaks and valleys within a row's $y$ coordinates to compute travel direction. First, the cart location data ($Y = \{(y_i)\}_{i=1}^N$) along the row length were smoothed using a Hampel filter. The Hampel filter is a specialized median filter that effectively detects and removes outlier data. The smoothed data were fed to the peak detection algorithm in \textit{SciPy} signal processing library. Based on the location and number of peaks and valleys, the picker travel direction was identified as -1 for the travel in decreasing y direction and +1 for travel in increasing y direction. If no peaks or valleys were found, the overall direction was computed based on the start and end harvest y coordinates. If one or more peaks and valleys were found, they were arranged alternately by removing consecutive peaks or valleys. After this step, a valley followed a peak, and vice versa. Then y-coordinates at the consecutive peak and valley locations $(i, i_{\text{next}})$ were used to estimate the travel direction for that specific row section.

\begin{algorithm}
\caption{Travel direction computation}
\label{alg:travel_dir}
\footnotesize
\begin{algorithmic}[1]
\Input{$Y = \{(y_i)\}_{i=1}^N$}
\Output{$dir$}

\Function{SIGN}{$y_1, y_2$}
    \State $\text{return } 1 \text{ if } y_1 - y_2 \geq 0, \text{ else } -1$
\EndFunction

\State $Y_{\text{filtered}} \gets \text{HampelFilter}(Y)$
\State $\text{idx\_pk}, \text{idx\_val} \gets \text{FindPeaksAndValleys}(Y_{\text{filtered}})$

\If{$\text{len}(\text{idx\_pk}) = 0$ and $\text{len}(\text{idx\_val}) = 0$}
    \State $dir \gets [\text{SIGN}(y_N, y_1)] * N$
\EndIf

\If{$\text{len}(\text{idx\_pk}) = 1$ and $\text{len}(\text{idx\_val}) = 0$}
    \State $dir \gets [\text{SIGN}(y_{\text{idx\_pk}_0}, y_1)] * \text{idx\_pk}_0 + [\text{SIGN}(y_N, y_{\text{idx\_pk}_0})] * (N - \text{idx\_pk}_0)$
\EndIf

\If{$\text{len}(\text{idx\_pk}) \geq 1$ \textbf{and} $\text{len}(\text{idx\_val}) \geq 1$}
    \State $\text{merged\_idx} \gets \text{AlternatePeakValleyMerge}(\text{idx\_pk}, \text{idx\_val})$
    \State $\text{merged\_idx} \gets [0] + \text{merged\_idx} + [N-1]$
    \For{\textbf{each} $(i, i_{\text{next}}) \in \text{merged\_idx}$}
        \State $dir \gets dir + [\text{SIGN}(y_{i_{\text{next}}}, y_i)] * (i_{\text{next}} - i)$
    \EndFor
\EndIf
\State \Return $dir$
\end{algorithmic}
\textbf{Note:} The `+' sign concatenates list elements together; the `*' sign is used to create a list of a specified length. For example, `[x] * n' creates a list with `n' repetitions of `x'.
\end{algorithm}

As illustrated in Algorithm \ref{alg:row_correction}, once the cart's travel direction was determined, it was assessed whether the cart was moving in the same direction across pairs and triplets of rows. If the travel direction matched among the pairs or triplets, this was regarded as a violation of the row completion condition. Among the rows where a condition violation was identified, the major row ($r_{\text{maj}}$), where the picker spent the most time, was considered the valid row in which the cart correctly performed harvesting. The row information from minor rows (those where the cart spent less time than the major row) was then reassigned to the major row. This was done by adding a new data column ($x_{\text{row\_new}}$) to the cart's data structure. A status flag column was added to each data point to indicate whether the row assignment had changed. The status flag for the cart location was set to 1 if the rows were reassigned; otherwise, it was set to 0. Figure \ref{fig:beforeafterrowcompletioncorrection}(Right) shows the corrected cart location traces after correcting the row completion condition violation.

\subsubsection{Step 4: Identification and resolution of row occupancy condition violation}
\begin{figure}
    \centering
    \includegraphics[width=0.15\linewidth]{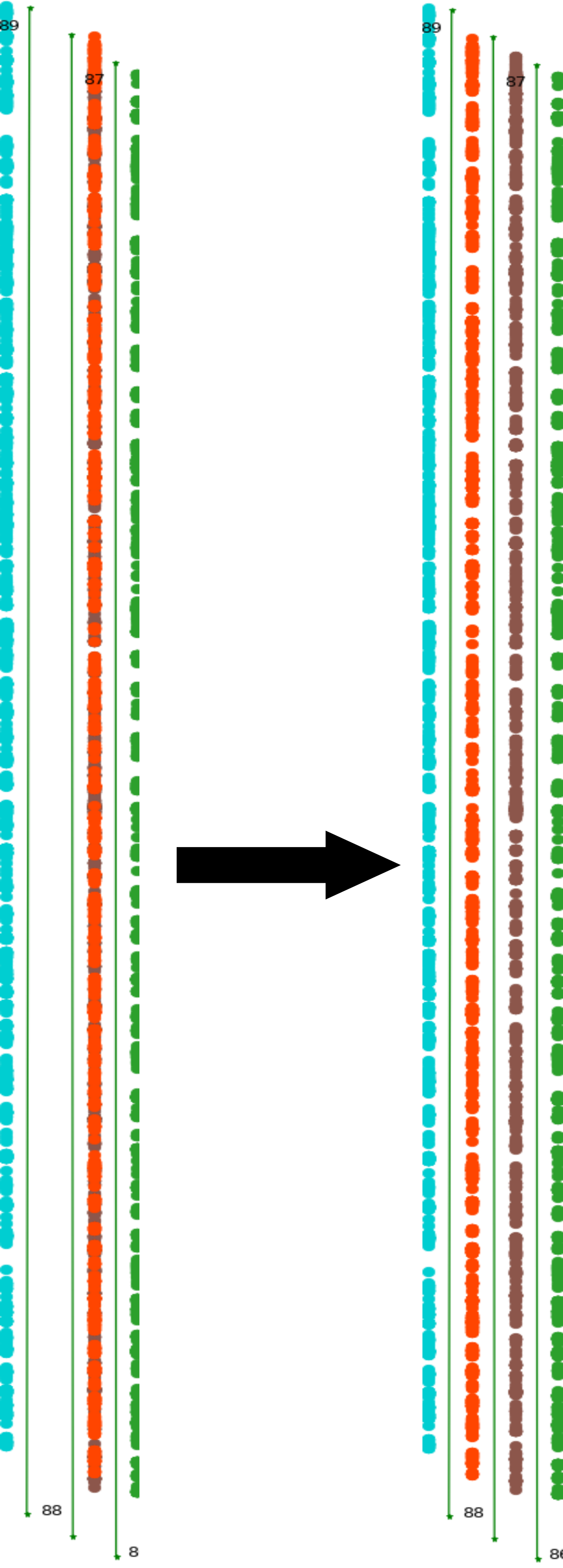}
    \caption{Mapped cart location traces before and after Step 4: Identification and resolution of row occupancy condition violation. (Left) Cart traces show multiple carts assigned to the same row, violating the row occupancy condition. The colored cart trace was incorrectly assigned to the neighboring row at the right with the purple cart trace. (Right) Corrected cart traces after resolving conflicts, with each row occupied by only one cart.}
    \label{fig:beforehand after row occupancy condition violation}
\end{figure}

The objective of this step was to identify and resolve the row occupancy condition violation such that only one cart was assigned to a harvest section of a row. If the GPS error persisted for an extended period, the cart's location may be recorded in a neighboring row for the majority of that harvest period. Since the resolution of the row completion condition violations depended on the amount of time the cart spent harvesting in a row, the algorithm may misinterpret the neighboring row as the one the cart was correctly harvesting. This leads to multiple carts appearing to harvest in a single row, resulting in a violation of row occupancy conditions (see Figure \ref{fig:beforehand after row occupancy condition violation} (Left)). To address this issue, data from all the carritos for an individual harvest day were combined into a table consisting of cart ID, location, timestamp, initially assigned row, and adjusted row and row adjustment flag added from Step 3 (Identification and resolution of row completion condition violation). The subset of carts ($C_r$) active in each row ($r \in R$) on a given harvest day was determined to identify the row occupancy condition violation. Let $(c_k, c_l) \in C_r \times C_r, k \neq l$ be a pair of carts harvesting in the same row $r \in R$. Next, the operational areas of these carts were examined for harvest location overlap. The overlap was calculated by using the y-coordinate ranges using the formula: $O_{kl} = \max(0, \min(y_{max}^k, y_{max}^l) - \max(y_{min}^k, y_{min}^l))$. If the calculated overlap \(O_{kl}\) was higher than a predetermined threshold, it was considered a violation of the row occupancy condition, and resolution steps were performed.

Algorithm \ref{alg:row_occupancy_correction} illustrates the details of the resolution of the row occupancy condition violation. Let $(c_k, c_l, r) \in \mathcal{C}$ represent the set of conflicting carts $c_k$ and $c_l$ in row $r$. First, the correction status flag for row completion violation was examined for both carts. The cart that was not previously reassigned (status flag = 0) during the row completion condition violation resolution was given priority to be in the conflicting row over the cart that had been reassigned (status flag = 1). The main idea was that if a cart had not previously been reassigned, it had already satisfied the row completion condition and had more confidence in its location data. If both cart rows were or were not previously adjusted during row occupancy condition, total time $T_k$ and $T_l$ spent by both carts $c_k$ and $c_l$ for harvesting was computed. The cart that spent most of the time harvesting in the conflicting row ($\max(T_k, T_l)$) was assigned to remain in row $r$. The remaining cart was reassigned to the row (left or right) from which the second majority of its data had originated ($r_{target}$) before the row occupancy correction was applied. Otherwise, the cart was reassigned to one of the empty neighboring rows ($r-1,r+1$) of the currently assigned row ($r$) or the neighboring rows ($r_{target}-1,r_{target}+1$) of second majority data row ($r_{target}$). Figure \ref{fig:beforehand after row occupancy condition violation} (Right) shows the corrected cart traces after resolving the row occupancy condition violation where each row was occupied by one cart.

\begin{algorithm}
\caption{Resolution of row occupancy condition violation}
\label{alg:row_occupancy_correction}
\footnotesize
\begin{algorithmic}[1]

\State \textbf{Input:} $D = \{(cart_{id_i}, x_i, y_i, t_i, x_{row_i}, x_{{row\_new}_i}, flag_i)\}_{i=1}^{NP}$, $\mathcal{C}$ (Set of conflicts), $R = \{(x_{r_j}, y_{r_j})\}_{j=1}^M$
\State \textbf{Output:} $C$ (Updated $C$ with resolved conflicts)

\Function{ReassignCart}{$cart, r_{new}$}
    \For{$c \in C$ where $c.cart_{id} = cart$}
        \State $c.x_{row\_new} \gets r_{new}.x_{r_j}$
    \EndFor
\EndFunction

\For{$(c_k, c_l, r) \in \mathcal{C}$}
    \State $adj_k \gets \textsc{HasBeenPreviouslyAdjusted}(c_k)$
    \State $adj_l \gets \textsc{HasBeenPreviouslyAdjusted}(c_l)$
    
    \If{$adj_k \neq adj_l$}
        \State $cart_{reassign} \gets \textbf{if } adj_k \textbf{ then } c_k \textbf{ else } c_l$
    \Else
        \State $T_k \gets \sum_{c \in C, c.cart_{id} = c_k, c.row_{id} = r} c.t$
        \State $T_l \gets \sum_{c \in C, c.cart_{id} = c_l, c.row_{id} = r} c.t$
        \State $cart_{reassign} \gets \textbf{if } T_k < T_l \textbf{ then } c_k \textbf{ else } c_l$
    \EndIf

    \State $r_{target} \gets \textsc{FindNextMajorDataRow}(cart_{reassign})$
    
    \State $candidates \gets [r_{target}, r-1, r+1, r_{target}-1, r_{target}+1]$
    
    \For{$r_{cand} \in candidates$}
        \If{$\textsc{IsRowAvailable}(r_{cand})$}
            \State \Call{ReassignCart}{$cart_{reassign}, r_{cand}$}
            \State \textbf{break}
        \EndIf
    \EndFor

\EndFor

\For{$cart \in \textsc{Set}(\{cart_{id_i}\})$}
    \State $C \gets C \cup \{d \mid d \in D, d.cart_{id} = cart\}$
\EndFor

\State \Return $C$

\end{algorithmic}
\end{algorithm}

\subsubsection{Step 5: Mass data processing}
\label{sec:mass_processing}
\begin{figure}
    \centering
    \includegraphics[width=0.48\linewidth]{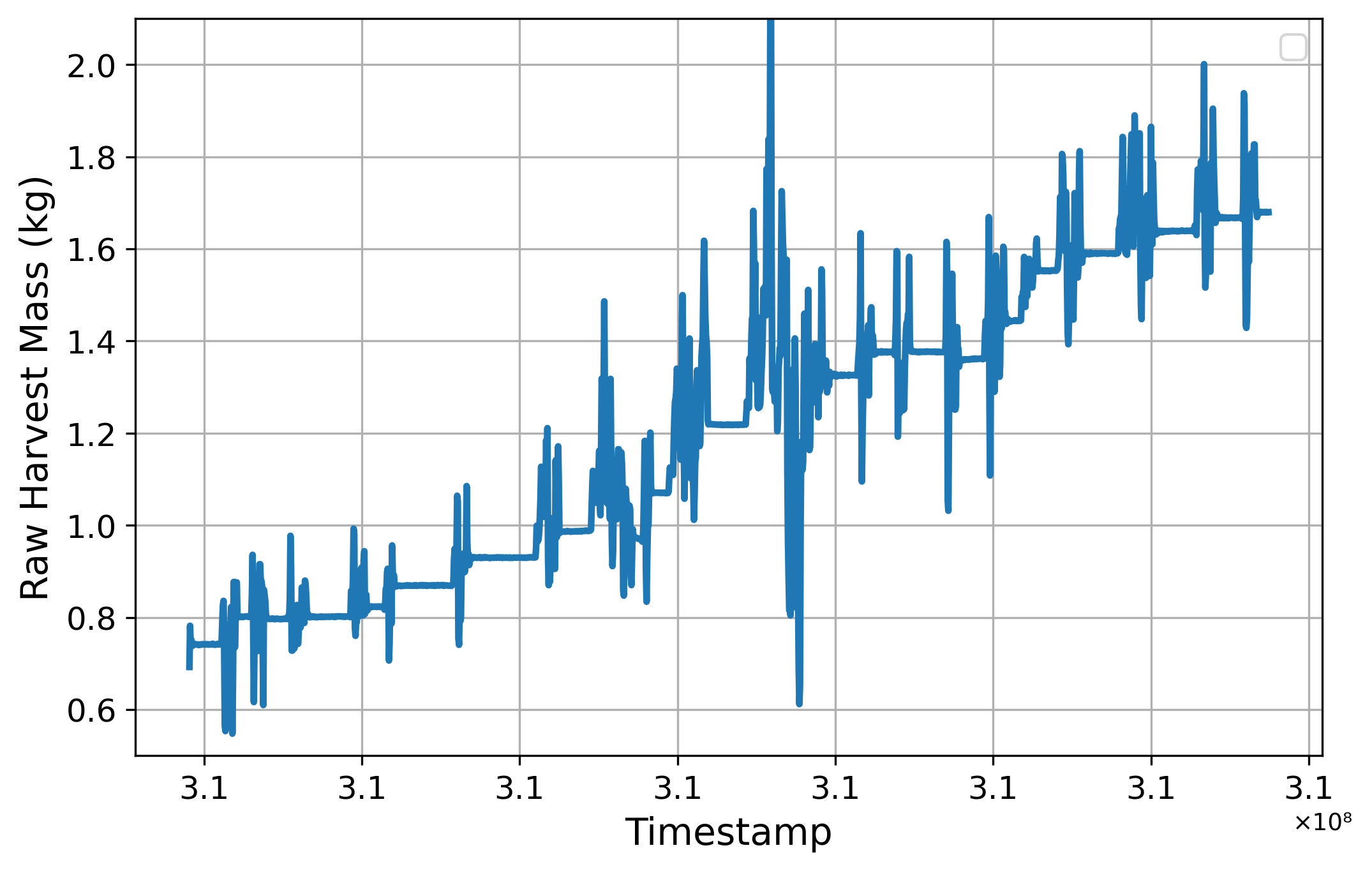}
    \includegraphics[width=0.48\linewidth]{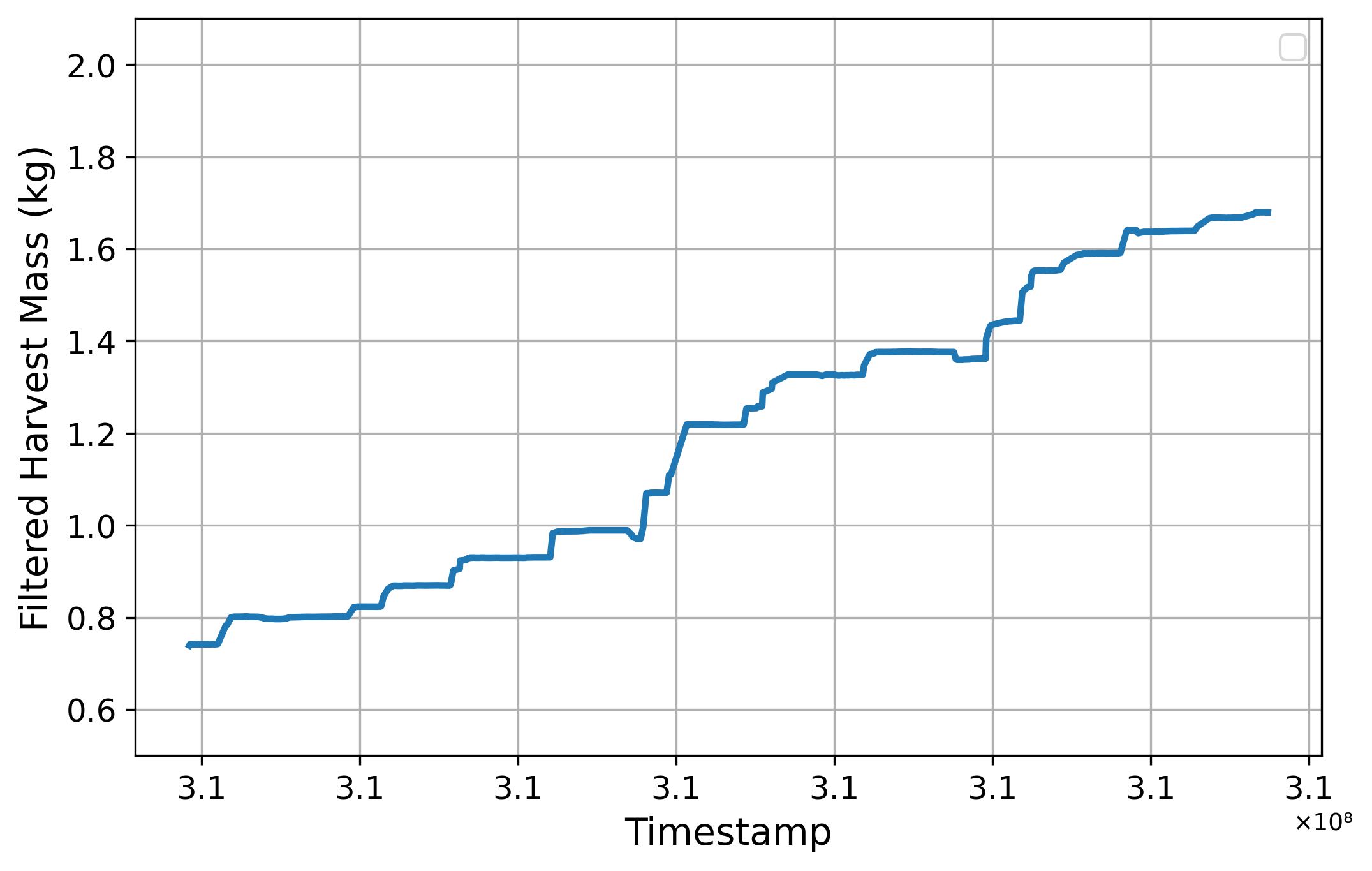}
    \caption{Harvest mass before and after Step 5: Mass data processing. (Left)Raw harvest mass showing a noisy staircase signal structure. (Right) The processed harvest mass displays a clear increasing trend of mass data.}
    \label{fig:staircasemass}
\end{figure}

The objective of this step was to clean up the load cell data by filtering out noise and retaining only valid measurements. While picking the berries, the pickers stop at a location, harvest the berries within reach, move a short distance, and continue harvesting in the same pattern (stop-harvest-move). Ideally, the mass was expected to increase as a staircase signal as the pickers harvested and added berries to the tray. However, various factors introduced noise into the mass data, such as sharp spikes that appeared when the picker placed berries in the tray, and large random forces applied to the load cells when the picker adjusted the tray on the carrito or pushed it (see Figure \ref{fig:staircasemass} (Left)). Algorithm \ref{alg:process_weight_data} was developed to retain stable incrementing mass while filtering out irrelevant weight spikes. First, the mass values outside the valid mass range $W_{\text{min}} < W < W_{\text{max}}$ with were removed. The range was defined based on average empty (0.55 kg) and full tray mass (5 kg). Next, the rate of change of harvest mass ($\dot{w}_i$) was computed to extract stable cart data ($C_{\text{stable}}$). The rate of change of mass was considered stable if it was below the maximum mass change threshold ($\Delta W_{\text{max}}$). Finally, the mass data was smoothed with a median filter to remove any remaining transient fluctuations and spurious noise. Figure \ref{fig:staircasemass} (Right) shows a clear trend of cumulative mass increments during harvesting. 

\begin{algorithm}
\caption{Process mass data}
\label{alg:process_weight_data}
\footnotesize
\begin{algorithmic}[1]
\Input{$C = \{(w_i, t_i)\}_{i=1}^N, \Delta W_{\text{max}}, w_{\text{max}}, w_{\text{min}}$}
\Output{$C_{\text{filtered}}$}

\State $W \gets \{w_i\}_{i=1}^N$
\State $W_{\text{range}} \gets W_{\text{min}} < W < W_{\text{max}}$
\State $C_{\text{valid}} \gets \{(w_j, t_j) \in C \mid W_{\text{range}}(w_j)\}_{j=1}^M$, where $M \leq N$
\State $W_{\text{valid}} \gets \{w_j\}_{j=1}^M$ from $C_{\text{valid}}$

\State $\dot{w}_j \gets \frac{w_{j+1} - w_j}{t_{j+1} - t_j}$ for $j = 1, \ldots, M-1$
\State $\text{stable\_idxs} \gets \{j \mid \dot{w}_j \leq \Delta W_{\text{max}} \land \dot{w}_j > 0\}$
\State $C_{\text{stable}} \gets \{(w_j, t_j) \in C_{\text{valid}} \mid j \in \text{stable\_idxs}\}$

\State $C_{\text{filtered}} \gets \text{MedianFilter}(\{w_j \mid (w_j, t_j) \in C_{\text{stable}}\})$

\State \Return $C_{\text{filtered}}$
\end{algorithmic}
\end{algorithm}

\subsubsection{Step 6: Yield estimation and mapping}
\begin{figure}
    \centering
    \includegraphics[width=0.3\linewidth]{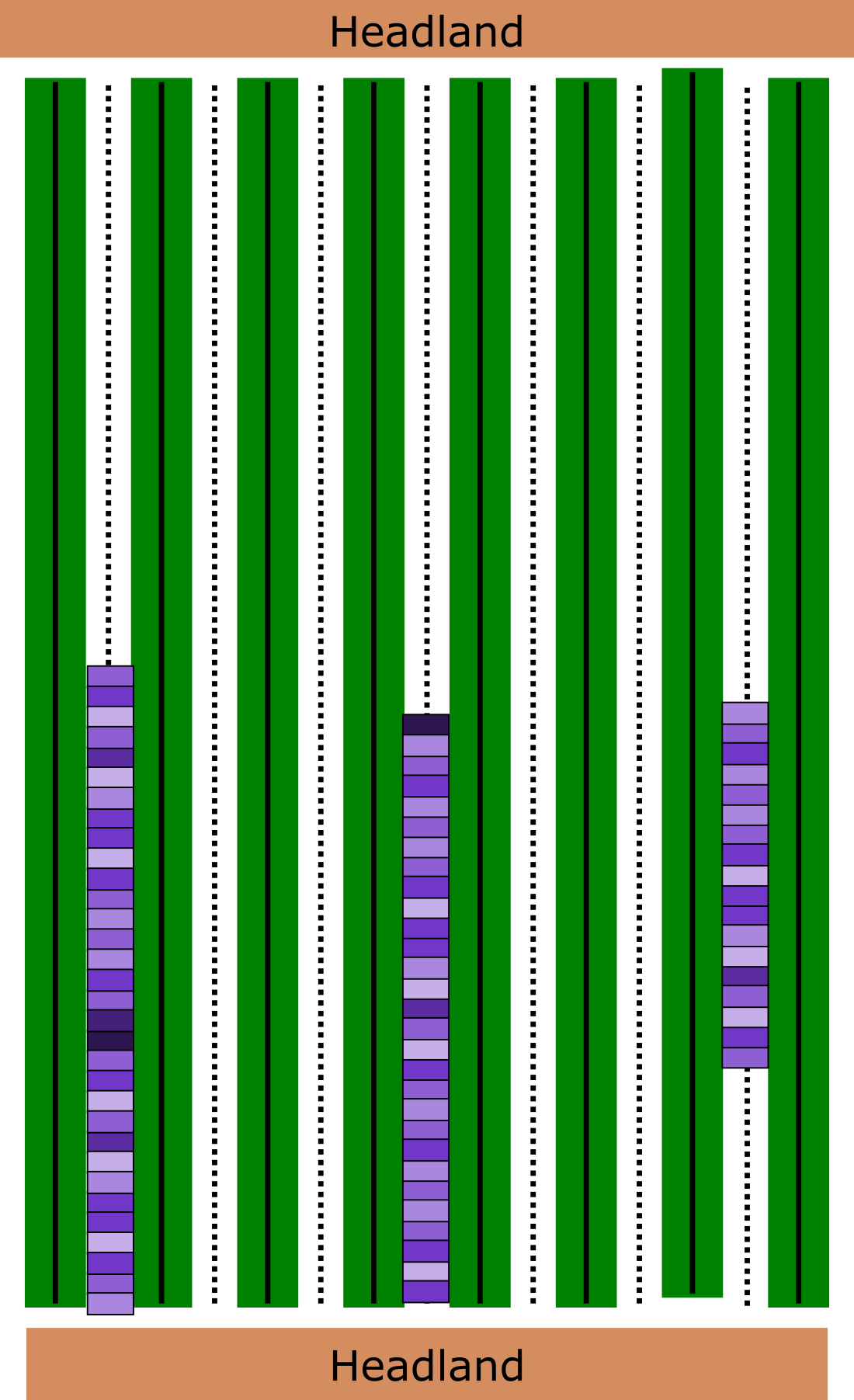}
    \caption{Step 6.1: Yield distribution estimation. Visual representation of interpolated mass data at a section of the field. The purple bands show estimated yield at uniform intervals of one foot.}
    \label{fig:interpolated yield computation one foot}
\end{figure}

\textbf{3.3.6.1 Step 6.1: Yield distribution estimation} \\
 \\
\label{sec: prelim_yield}
As discussed in Section \ref{sec:mass_processing}, pickers do not place strawberries in the cart or push the cart at regular time or space intervals. Sometimes, a picker may walk for several feet, placing strawberries in a clamshell and only putting the clamshell in the cart once it is full. In such scenarios, there will be no change in the load cell signal for some distance, although berries were harvested as the worker was traversing this distance. To address this issue, yield distribution inside a line segment at a small, incremental distance was estimated using mass data interpolation along y coordinates. The data was interpolated using adaptive polynomial regression, and the harvest mass was computed for the incremental distance of one foot as shown in Figure \ref{fig:interpolated yield computation one foot}. The sampling interval of 0.3048 meters (1 ft) was selected since the strawberry plants are planted at a spacing of 0.3048 meters (1 ft) to 0.4572 meters (1.5 ft) apart to allow proper airflow and growth \citep{vossen}.

\begin{algorithm}
\caption{Separate each tray for data interpolation}
\label{alg:assign_tray_ids}
\footnotesize
\begin{algorithmic}[1]
\Input{$C = \{(y_i, t_i, w_i, x_{\text{row\_new}_i})\}_{i=1}^N$}
\Output{Updated $C$ with appended $\text{tray}_{id}$}

\State $labels \gets \text{DBSCAN}(C, \epsilon, minPts)$
\For{$tray_{id}$ in $labels$ \textbf{where} $tray_{id} \neq noise_{id}$}
        \For{$point$ in $tray_{id}$} 
            \State $C.append(tray_{id})$
        \EndFor
\EndFor
\State \Return $C$
\end{algorithmic}
\end{algorithm}

For interpolation, the data associated with a full tray or partially full tray were extracted first. Algorithm \ref{alg:assign_tray_ids} illustrates the process of clustering the extracting data and assigning them with a unique identifier ($tray_{id}$). Given the set harvest data $C = \{(y_i, t_i, w_i, x_{\text{row\_new}_i})\}_{i=1}^N$, DBSCAN algorithm was used to assign harvest data to a unique cluster or classify as noise. Once all the cart data were clustered and assigned with a unique identifier ($tray_{id}$), a data interpolation was done for each unique data cluster. Algorithm \ref{alg:yield_fine_res} details the process of interpolated yield computation at the sampling interval of 0.3048 meters (1 ft) using adaptive polynomial regression. First, the harvest data $(C_{row})$ was extracted for each unique row $(x_{row\_new})$, and tray ID $(tray_{id})$.  Next, the best polynomial (from first to third order) was determined to fit the filtered mass data best. A polynomial was considered the best fit if its fit score was more than 0.94. The selected polynomial was then used to predict mass at y-coordinates spaced apart by 0.3048 meters ($Y_{int}$), and absolute differences between consecutive predicted mass (\(\Delta W\)) were used to estimate the yield.

\begin{algorithm}
\caption{Yield interpolation using adaptive polynomial regression}
\label{alg:yield_fine_res}
\begin{algorithmic}[1]
\Input{$C = \{(x_i, y_i, t_i, w_i, x_{{row\_new}_i}, tray_{id_i})\}_{i=1}^N$, $interval = 0.3048$, $score_{max}$, $W_{\text{max}}$, $R = {(x_{r_j}, y_{r_j})}_{j=1}^M$}

\Output{$Y$ (Yield data)}
\State $Y \gets \emptyset$
\For{each unique $(tray_{id}, x_{row\_new})$ in $C$}
    \State $C_{row} \gets \{c \in C | c.tray_{id} = tray_{id}, c.x_{row\_new} = x_{row\_new}\}$
        
        \State $opt\_model, opt\_degree \gets \text{FitBestPolynomial}(C_{row}.y, C_{row}.w, score_{max},deg_{max}=3)$
        \State $Y_{int} \gets \{y_{int} | y_{min} \leq y \leq y_{max}, \text{step} = interval\}$
        \State $W_{pred} \gets \text{PredictMass}(opt\_model, Y_{int}, opt\_degree)$
        \State $\Delta W \gets |\text{diff}(W_{pred})|$
        \State $Y \gets Y \cup \{(C_{row}.x_{row\_new}, y_{int}, \delta w_{int}) | y_{int} \in Y_{int}, \delta w_{int} \in \Delta W\}$

\EndFor
\State \Return $Y$
\end{algorithmic}
\end{algorithm}

\textbf{3.3.6.2 Step 6.2: Yield map generation at a desired resolution} \\
This step aimed to create a yield map at a desired grid resolution. First, the entire field was divided into a grid with cells of the desired size, and then the interpolated yield data was assigned to and summed up within those grid cells. Figure \ref{fig:yield at desired resolution} (left) illustrates the visual representation of the yield map generation process. As pickers pick berries from two beds on the left and right sides of the row, the grid width was selected to match the average distance between two neighboring bed centers. The grid length could vary along the row. Given that the row spacing in Santa Maria was 163 cm and in Salinas was 122 cm, a square grid with each cell side equal to the row spacing was selected for this work.

Algorithm \ref{alg:yield_desired_res} illustrates the process of generating a yield map at resolution ($r$). Let $Y=\{x_{{row\_new}_i}, y_{{int}_i}, \delta w_{{int}_i}\}_{i=1}^N$ be the interpolated yield data and $\{[x_{min}, y_{min}], [x_{max}, y_{max}]\}$ be the lower left and upper right field corner coordinates. First, the field was divided into grids of the size defined by the resolution, and the center of each grid cell was computed. Next, each yield distribution data \((x_{{row\_new}_i}, y_{{int}_i}, \delta w_{{int}_i})\) computed from Step 6.1 was then assigned to the closest grid cell based on the absolute distance along x and y direction from the grid center. Finally, the cumulative yield for each grid cell was computed by summing all assigned yield data points from all carritos to each grid cell. Figure \ref{fig:yield at desired resolution} (Right) shows the generated yield map for a single harvest day during a typical strawberry harvest in Santa Maria. Once the yield map for each harvest day was generated, the yield map for the entire season or the multiple harvest days was generated by accumulating the yield at each grid cell during the harvest season.

\begin{algorithm}
\caption{Compute Mapping for Desired Field Resolution}
\label{alg:yield_desired_res}
\begin{algorithmic}[1]
\Input{$Y=\{x_{{row\_new}_i}, y_{{int}_i}, \delta w_{{int}_i}\}_{i=1}^N$, $\{[x_{min}, y_{min}], [x_{max}, y_{max}]\}$, $r$}
\Output{$Y_r$}
\State $x_{edges}, y_{edges} = [x_{min}, x_{min} + r, ..., x_{max}], [y_{min}, y_{min} + r, ..., y_{max}]$

\State $x_{mid,j} = (x_{edges,j} + x_{edges,j+1}) / 2, \quad \text{for } j = 1,\ldots,m-1$
\State $y_{mid,k} = (y_{edges,k} + y_{edges,k+1}) / 2, \quad \text{for } k = 1,\ldots,n-1$
\For{each yield data point $(x_{{row\_new}_i}, y_{{int}_i}, \delta w_{{int}_i}) \text{ in } Y$}
    \State $Y_{idx\_x,i} \gets \argmin_j |x_{mid,j} - x_{{row\_new}_i}|$
    \State $Y_{idx\_y,i} \gets \argmin_k |y_{mid,k} - y_{{int}_i}|$
\EndFor
\For{each grid cell $(j, k)$}
    \State $cell\_idx \gets \{i : Y_{idx\_x,i} = j \text{ and } Y_{idx\_y,i} = k\}$
    \State $Y_r[j, k] \gets \sum_{i \in cell\_idx} \delta w_{{int}_i}$
\EndFor
\State \Return $Y_r$
\end{algorithmic}
\end{algorithm}

\subsection{Experimental Design for Performance Evaluation}

\subsubsection{Evaluation of yield distribution estimation}
\label{sec:evaluation of yield distribution}

\begin{figure}
    \centering
    \includegraphics[width=0.4\linewidth]{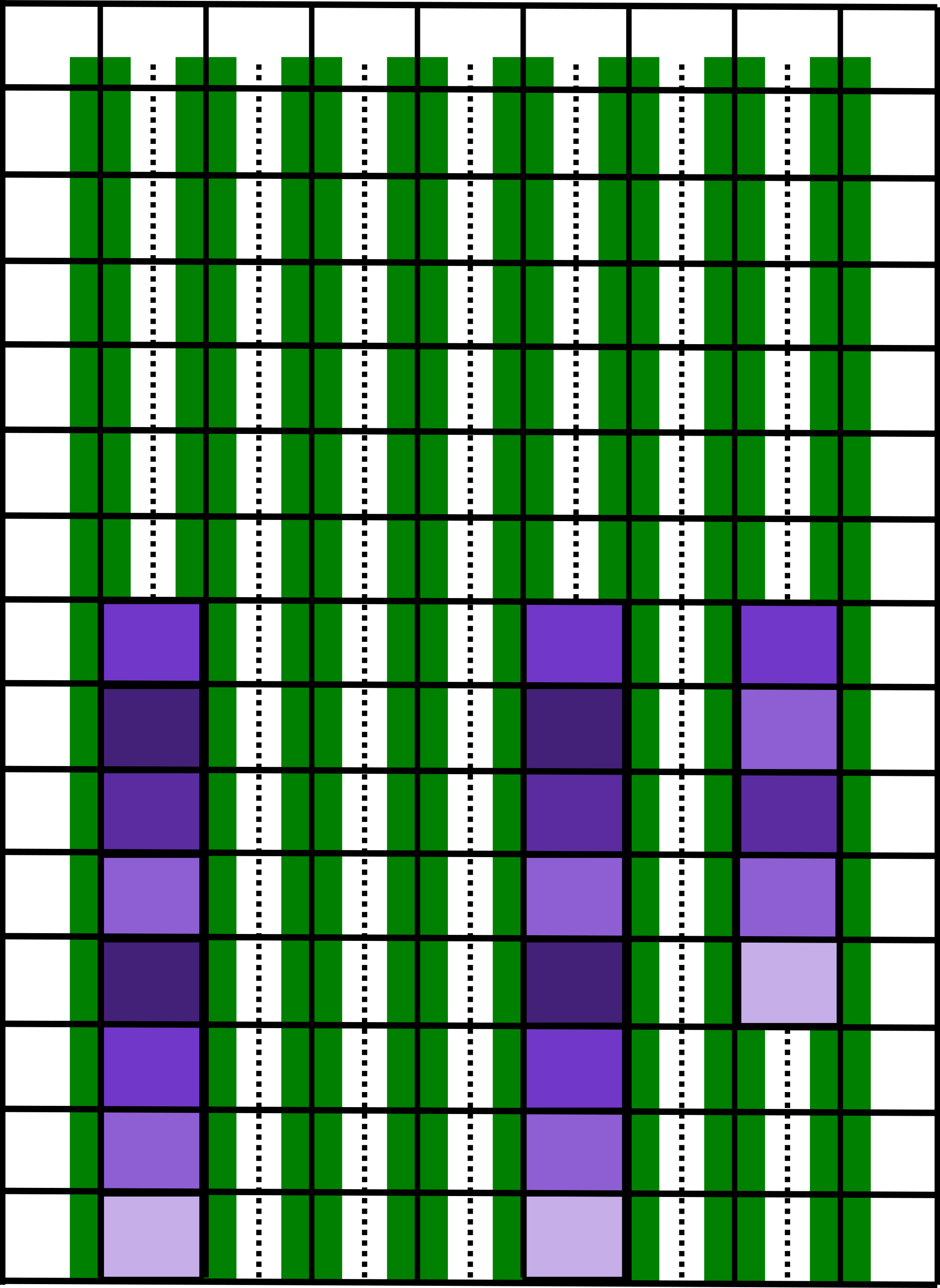}
    \includegraphics[width=0.57\linewidth]{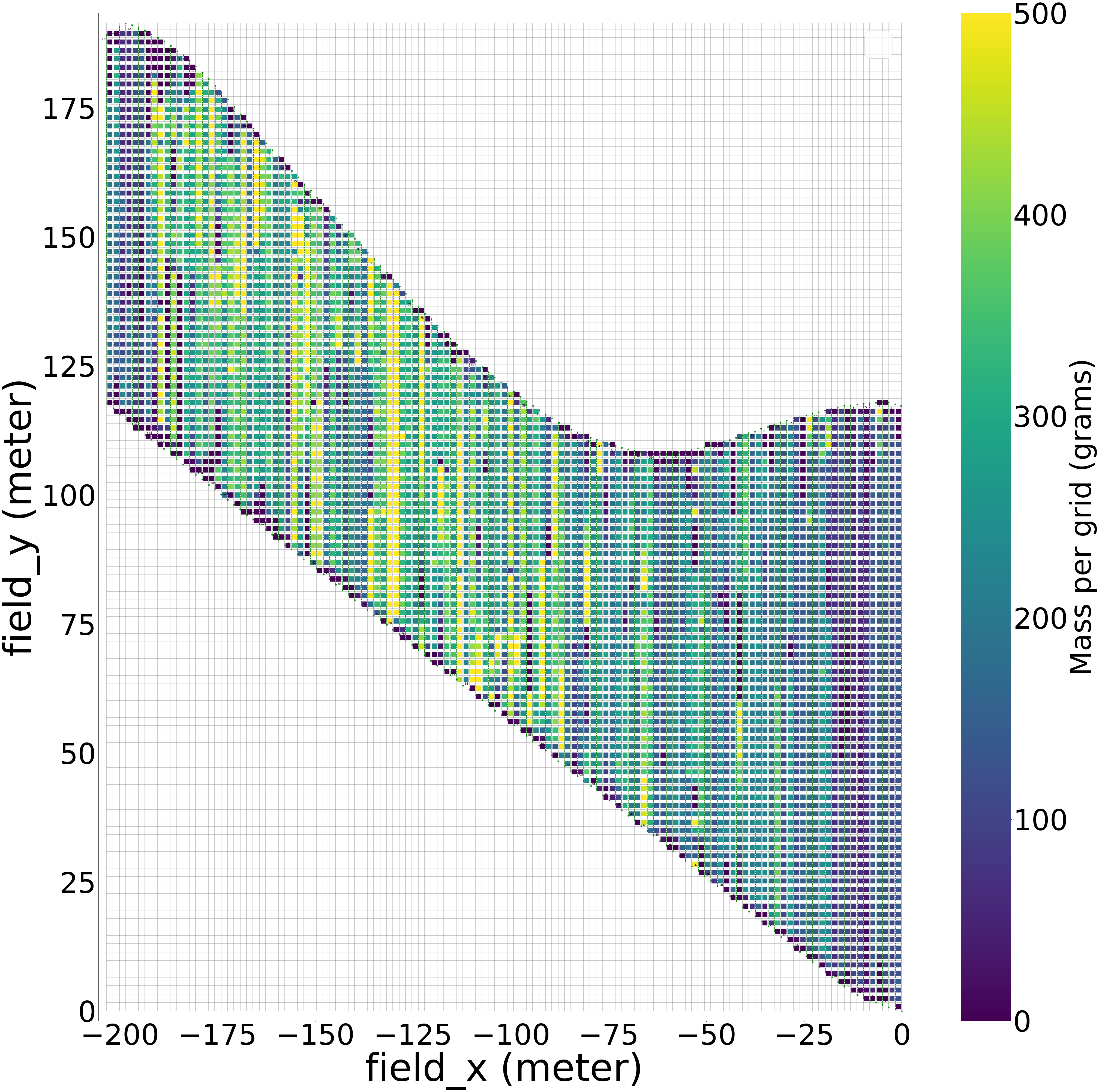}
    \caption{Step 6.2: Yield map generation at a desired resolution. (Left) Schematic representation of yield map generation at square grid size equals row spacing. The purple bands show accumulated yield data in each grid. (Right) Generated yield map for a single harvest day during a typical strawberry harvest in Santa Maria. The color scale represents mass per grid in grams, with darker blue indicating lower yield and brighter yellow indicating higher yield.}
    \label{fig:yield at desired resolution}
\end{figure}

\begin{figure}
    \centering
    \includegraphics[width=0.32\linewidth]{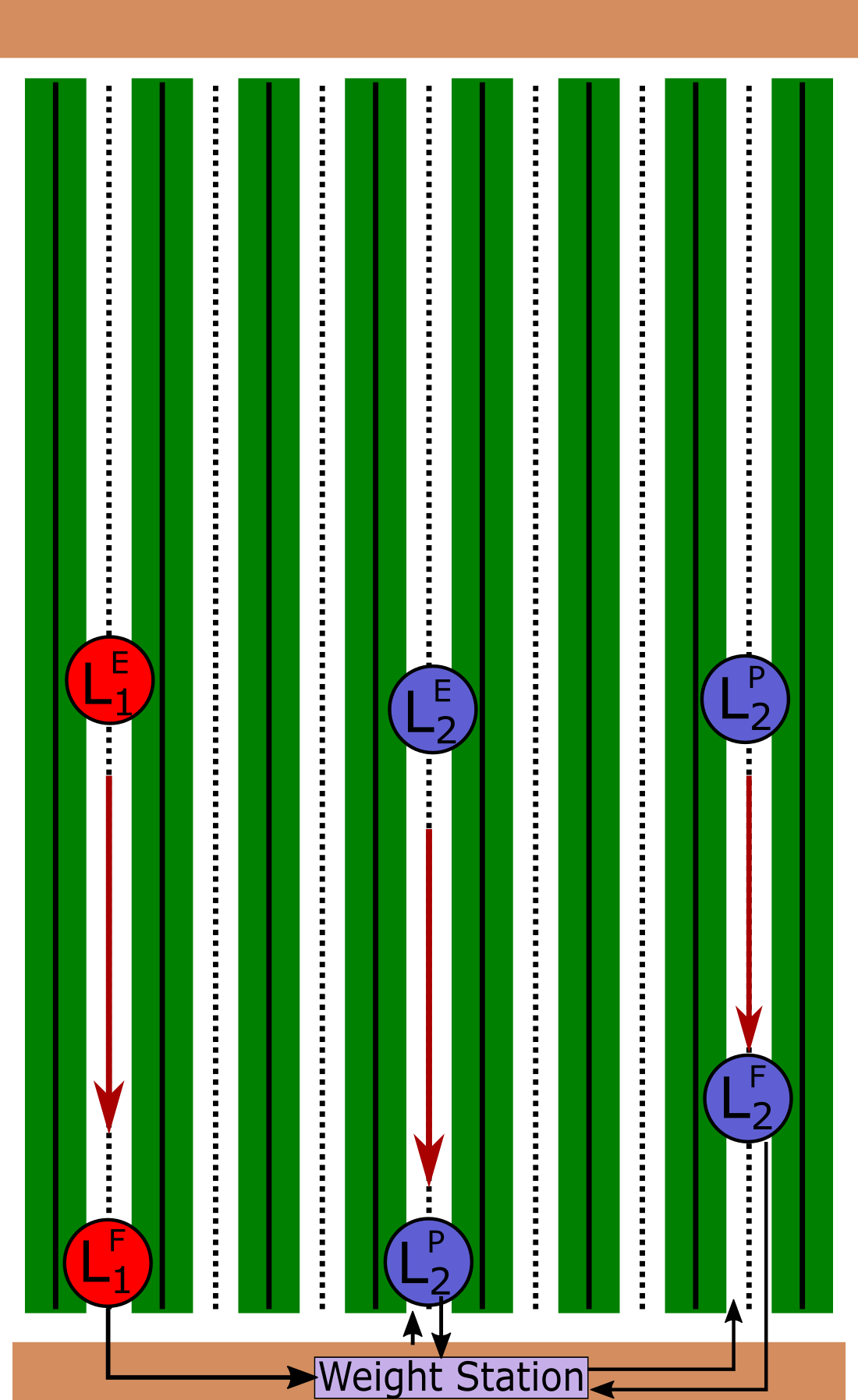}
    \caption{Schematic representation of strawberry harvesting during ground truth data collection. Green strips represent the plant beds, and red arrows show the direction of picker movement. Red and blue circles indicate different trays and their start and end harvest positions in a row. Tray status is represented by letters E (empty), F (full), and P (partially full). The mass of the full tray was measured during delivery, while the mass of the partially full tray was recorded before switching to a new row from the current harvest row.}
    \label{fig:gt_data_collection}
\end{figure}

To evaluate the accuracy of the yield distribution estimation, ground truth data was collected using the iCarritos during commercial strawberry harvest in Salinas, CA. A total of five pickers were randomly selected and observed for one hour. Each picker was accompanied by two people: one person recorded the timestamps of the different states as defined by equation \ref{eq:pickerstates}, while the other placed flags to mark the locations of key events: an empty or partially full tray was placed on the iCarrito, a full tray was lifted from the iCarrito for delivery to the collection station, or a partially full tray was lifted from the iCarrito to continue harvesting in another row. Figure \ref{fig:gt_data_collection} illustrates the process for a picker harvesting two trays (identified by red and blue colors), with key data collection locations marked by $L_{\text{tray\_id}}^{\text{tray\_status}}$. The tray\_status was categorized as empty (E), full (F), and partially full (P). Furthermore, each time a tray was lifted from the iCarrito, its mass was measured by a third person at the weigh station on the headland. Figure \ref{fig:gt_data_collection} shows that the first tray was full while harvesting in a single row, and its mass was measured during delivery. To fill the second tray, the picker had to visit two rows. Hence, the tray mass was measured twice: first, when it was partially full before switching to the new row, and then when it was full and being delivered. The average net mass of the harvested trays was 4.25 Kg (after subtracting the tray mass of 0.55 Kg). Each picker's carrito data during the one-hour observation was processed using the yield estimation pipeline from Step 1 to Step 6.1 to generate the yield distribution of the rows harvested in that period. The reliability of the yield distribution estimation was evaluated at two levels: (1) the row segment level, where the estimated mass of berries harvested in each row segment was compared to ground truth, and (2) the full tray level, where the estimated full tray mass was compared to ground truth. 

To evaluate the yield distribution estimation on a row segment, the total mass of strawberries harvested in each row segment for each tray was calculated by summing all masses from the generated yield distribution map. Based on the illustration in Figure \ref{fig:gt_data_collection}, the cumulative mass will be calculated from three distinct row segments: $L_1^E$ to $L_1^F$ for the first tray, $L_2^E$ to $L_2^P$ for the second tray's first-row segment, and $L_2^P$ to $L_2^F$ for the second tray's second-row segment. These calculated masses were then compared to the manually measured masses (ground truth). The accuracy was calculated as:

\begin{equation}
\text{Row level yield distribution estimation accuracy} = \left(1 - \frac{1}{T} \sum_{i=1}^T \sum_{j=1}^{R_i} \frac{|\text{gt\_mass}_{i,j} - \text{est\_mass}_{i,j}|}{\text{gt\_mass}_{i,j}}\right) \times 100\%
\end{equation}

Where $T$ was the total number of trays, $R_i$ was the number of rows harvested to fill tray i, $\text{gt\_mass}_{i,j}$ was ground truth mass from row j for tray i, and $\text{est\_mass}_{i,j}$ was estimated mass from row j for tray i.

To evaluate the yield distribution based on full tray mass, the full tray mass was computed by adding all masses from the row(s) that contributed to filling up a tray in the generated yield distribution map. Based on the illustration in Figure \ref{fig:gt_data_collection}, the cumulative mass will be calculated for two full trays: $L_1^E$ to $L_1^F$ for the first tray, $L_2^E$ to $L_2^F$ for the second tray. These calculated full tray masses were then compared to the manually measured ones. The accuracy was calculated as:

\begin{equation}
\text{Tray level yield distribution estimation accuracy} = \left(1 - \frac{1}{T} \sum_{i=1}^T \frac{|\sum_{j=1}^{R_i}\text{gt\_mass}_{i,j} - \sum_{j=1}^{R_i}\text{est\_mass}_{i,j}|}{\sum_{j=1}^{R_i}\text{gt\_mass}_{i,j}}\right) \times 100\%
\end{equation}

$T$ was the total number of trays, $R_i$ was the number of rows harvested to fill tray i, and $\sum_{j=1}^{R_i}$ computed cumulative mass from all rows used to fill tray i.

\subsubsection{Evaluation of yield estimation based on tray count}
To evaluate yield estimation for the entire harvest season, daily harvest data was first processed through the yield estimation pipeline to generate a yield distribution map. Then, the total harvest mass for each carrito was calculated by summing up the yield data from the generated yield distribution map. The resulting mass was then divided by the average full tray mass (4.25 kg) to compute the number of harvested trays for each iCarrito on each harvest day. The number of full trays was manually counted from the raw harvest data to generate the ground truth. The manual count was validated by cross-referencing it with the harvest tray counts provided by the grower to compensate the pickers. 

\begin{equation}
\text{Tray count estimation accuracy} = \left(1 - \frac{1}{C} \sum_{i=1}^C \sum_{j=1}^{D_i} \frac{|\text{gt\_count}_{i,j} - \text{est\_count}_{i,j}|}{\text{gt\_count}_{i,j}}\right) \times 100\%
\end{equation}

Where $C$ was the total number of iCarritos, $D_i$ was the number of harvest days for carrito i, $\text{gt\_count}_{i,j}$ was ground truth tray count for carrito i on day j, $\text{est\_count}_{i,j}$ was estimated tray count for carrito i on day j

\section{Results and Discussion}
\label{sec:resultsdiscussion}
\subsection{Evaluation of yield distribution estimation}
\begin{figure}[h!]
    \centering
    \includegraphics[width=0.51\linewidth]{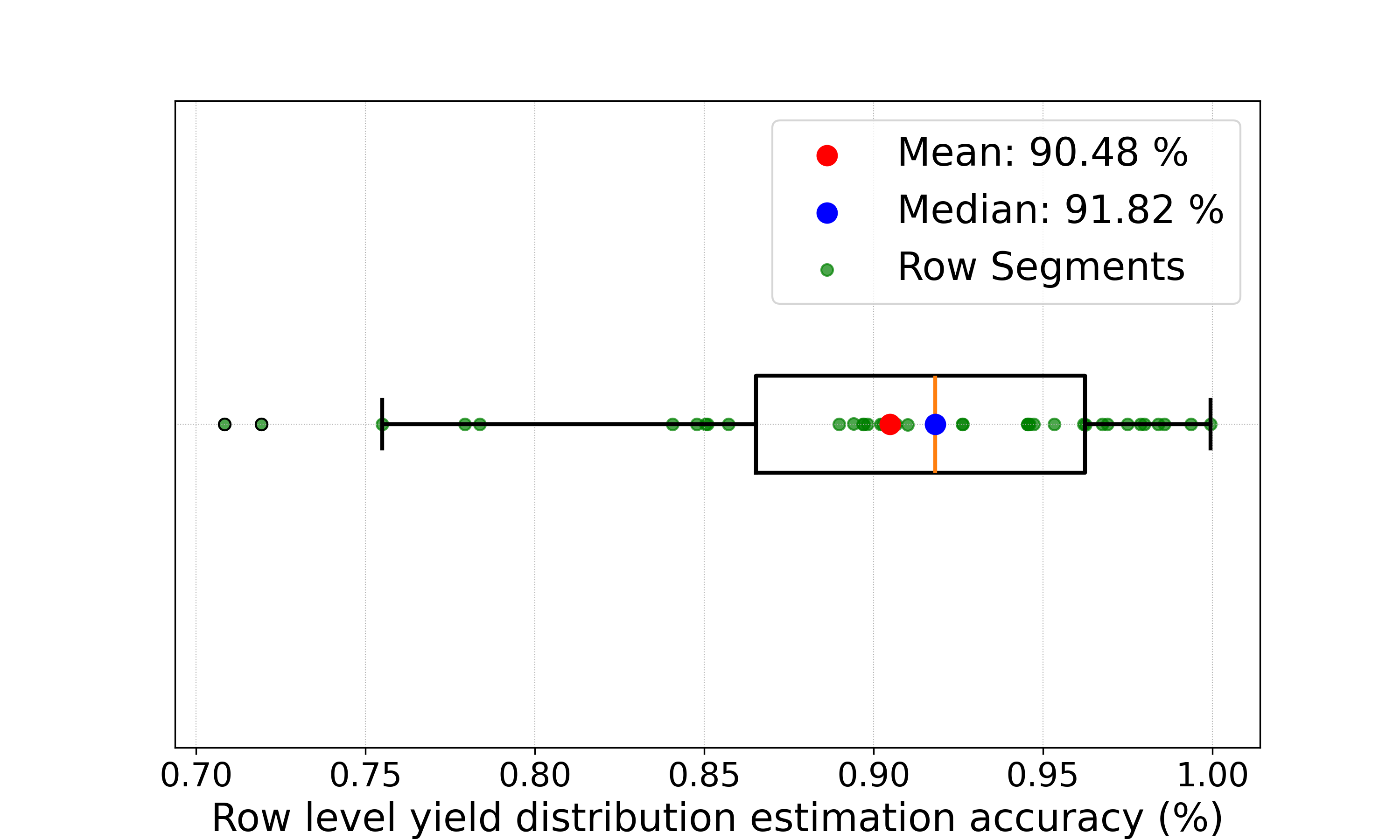}
    \includegraphics[width=0.48\linewidth]{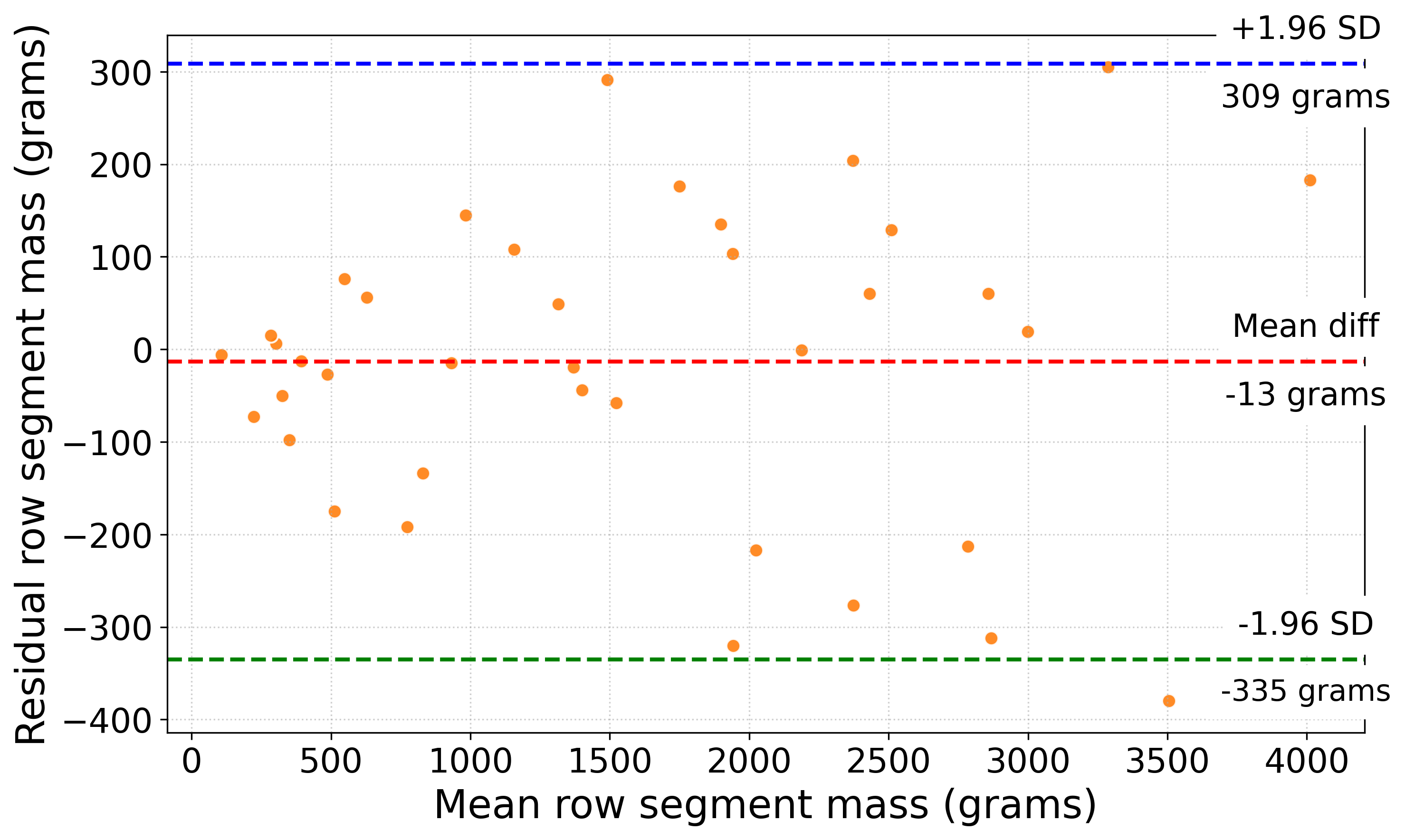}
    \caption{(Left) Box plot showing the distribution of row-level yield estimation accuracy with a mean of 90.48\% and median of 91.82\%. (Right) Bland-Altman plot comparing estimated and actual harvested mass, showing a mean underestimation bias of 13 grams with 95\% limits of agreement at $\pm$335 grams.}
    \label{fig:yield_segment_weight_accuracy}
\end{figure}
Figure \ref{fig:yield_segment_weight_accuracy} (left) shows a box plot illustrating the yield distribution estimation accuracy at the row segment level. The mean and median accuracy were 90.48\% and  91.82\%, respectively, with accuracy scores ranging from 70.9\% to 99.8\%. Additionally, for tray-level yield distribution estimation accuracy, the proposed pipeline achieved a mean accuracy of 94.05\%, with accuracy ranging from 89.71\% to 99.15\% and low variability in estimation indicated by a standard deviation of 2.79\%. A Bland-Altman analysis was performed to further evaluate estimated versus actual mass at the row segment level. This involved plotting the residuals of the ground truth against the estimated mass versus the mean harvest mass, as shown in Figure \ref{fig:yield_segment_weight_accuracy} (right). The analysis showed the yield distribution estimation pipeline had a minor bias, underestimating the harvested mass by an average of 13 grams. Furthermore, for 95\% of the harvest data, the estimated harvest mass was likely to fall within 335 grams of the actual ground truth. Notably, in 2024, the Salinas field suffered from low yield due to a disease outbreak, which led growers to stop harvesting mid-season. This is reflected in the Bland-Altman plot, where only one tray was nearly fully harvested from a single row. Due to the sparse yield, pickers had to switch between multiple rows to fill a tray. Some rows recorded very low yields, around 300 grams, which caused small variations in mass estimation to appear as high percentage errors. This situation is not typical. Under normal circumstances, multiple trays are harvested from a single row.

\subsection{Evaluation of yield estimation and yield map generation}

\subsubsection{Evaluation of yield estimation based on tray count}
The proposed yield estimation system demonstrated a promising accuracy of more than 94\% in estimating yield based on tray count. Despite the variability across different fields, strawberry varieties, and planting structures, the system's accuracy was high and consistent. Table \ref{tab:results_comparison} summarizes the evaluation of the tray count from the proposed yield estimation pipeline in comparison to the ground truth. For Santa Maria, the average tray count accuracy per cart was 94.89\%, with a mean absolute error (MAE) of 1.70 trays. Out of a total of 14282 harvested trays, 14217 trays were estimated through the pipeline. Similarly, for Salinas, the average tray count accuracy per cart was 94.20\%, with an MAE of 1.72 trays. Out of 11254 total harvested trays in Salinas, the estimated tray count from the pipeline was 10959.

\begin{table}[ht]
\small
\centering
\caption{Comparison of ground truth and the estimated tray count for all the harvest days}
\begin{tabular}{lcc}
\hline
\textbf{Metric}               & \textbf{Santa Maria} & \textbf{Salinas} \\ \hline
Total harvested trays           &14282                & 11254            \\
Total estimated trays           &14217                & 10959            \\
Mean tray count accuracy per cart                   & 94.89\%              & 94.20\%          \\ 
Mean Absolute Error (MAE)      & 1.70                 & 1.72             \\ 
Root Mean Squared Error (RMSE) & 2.71                 & 2.60             \\ 
Pearson r                      & 0.99                 & 0.99             \\ \hline
\end{tabular}
\label{tab:results_comparison}
\end{table}

\begin{figure}[ht]
    \centering
        \includegraphics[width=0.48\textwidth]{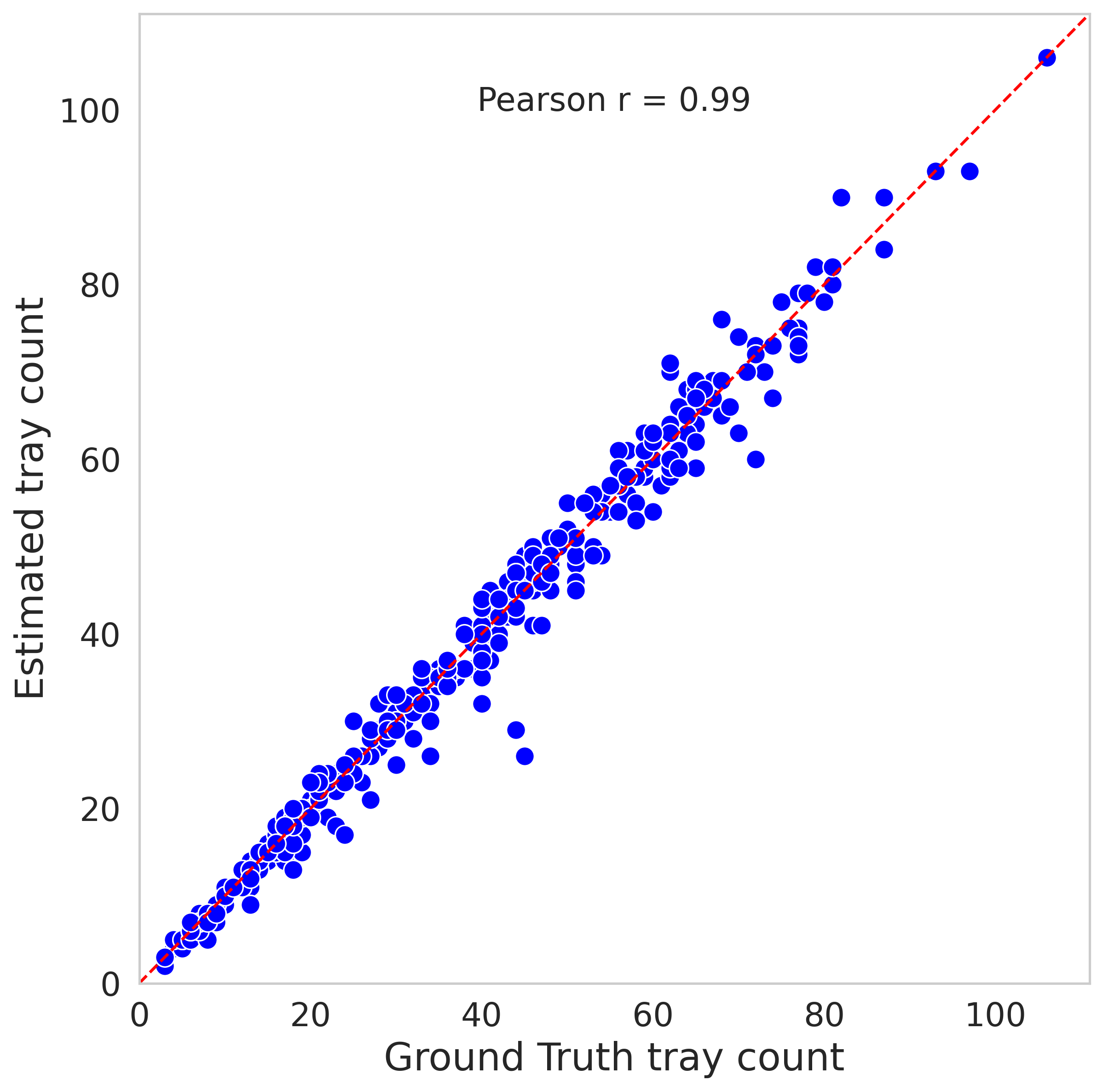}
        \includegraphics[width=0.48\textwidth]{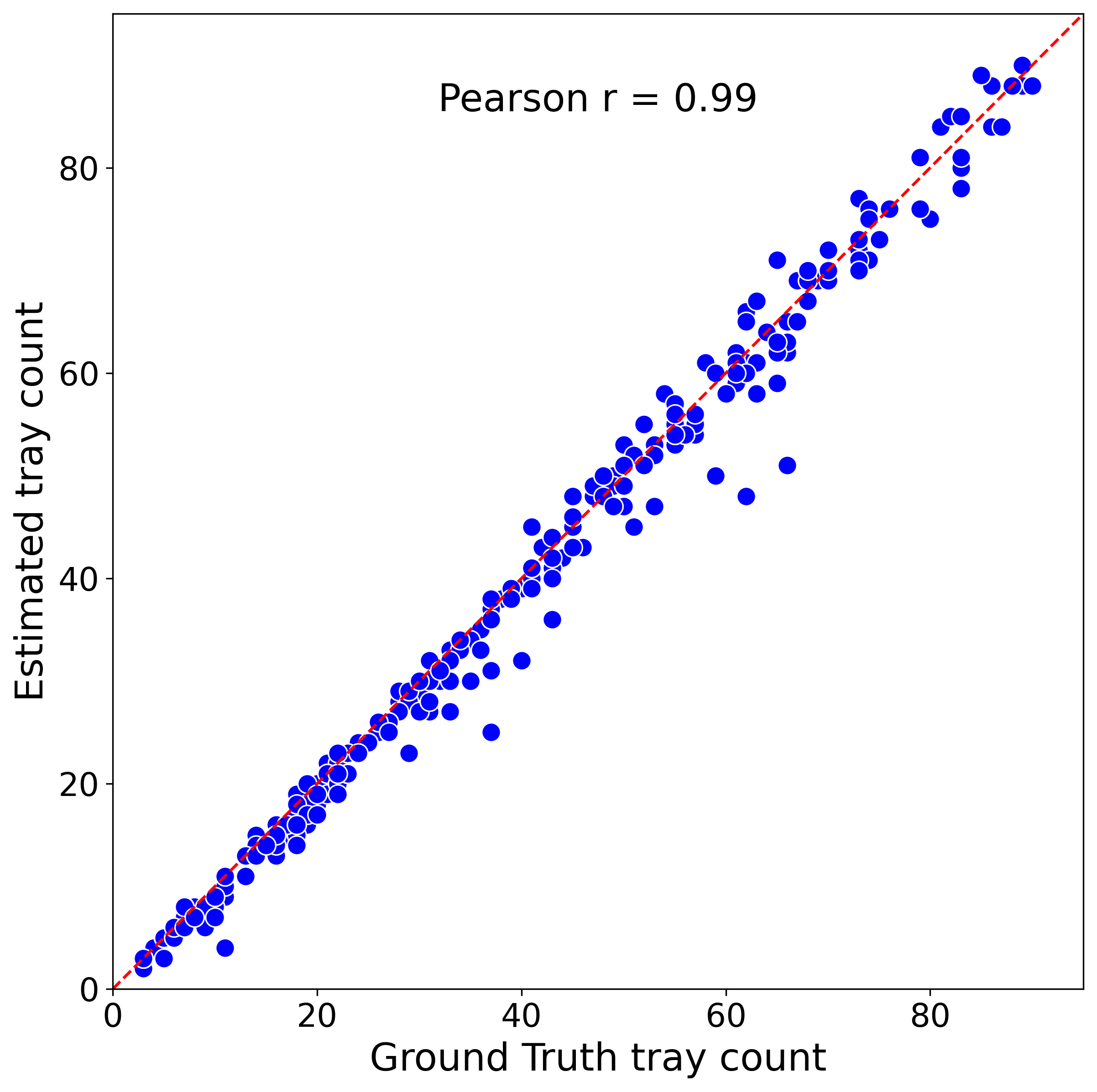}

    \caption{Scatter plots showing a strong correlation (Pearson r = 0.99) between estimated tray count vs ground truth tray count for each cart in Santa Maria (Left) and Salinas (Right) for all harvest days. The red dashed line represents the ideal fit between the estimated and actual tray count.}
    \label{fig:scatter_plots_comparison}
\end{figure}

\begin{figure}[ht]
        \includegraphics[width=0.48\textwidth]{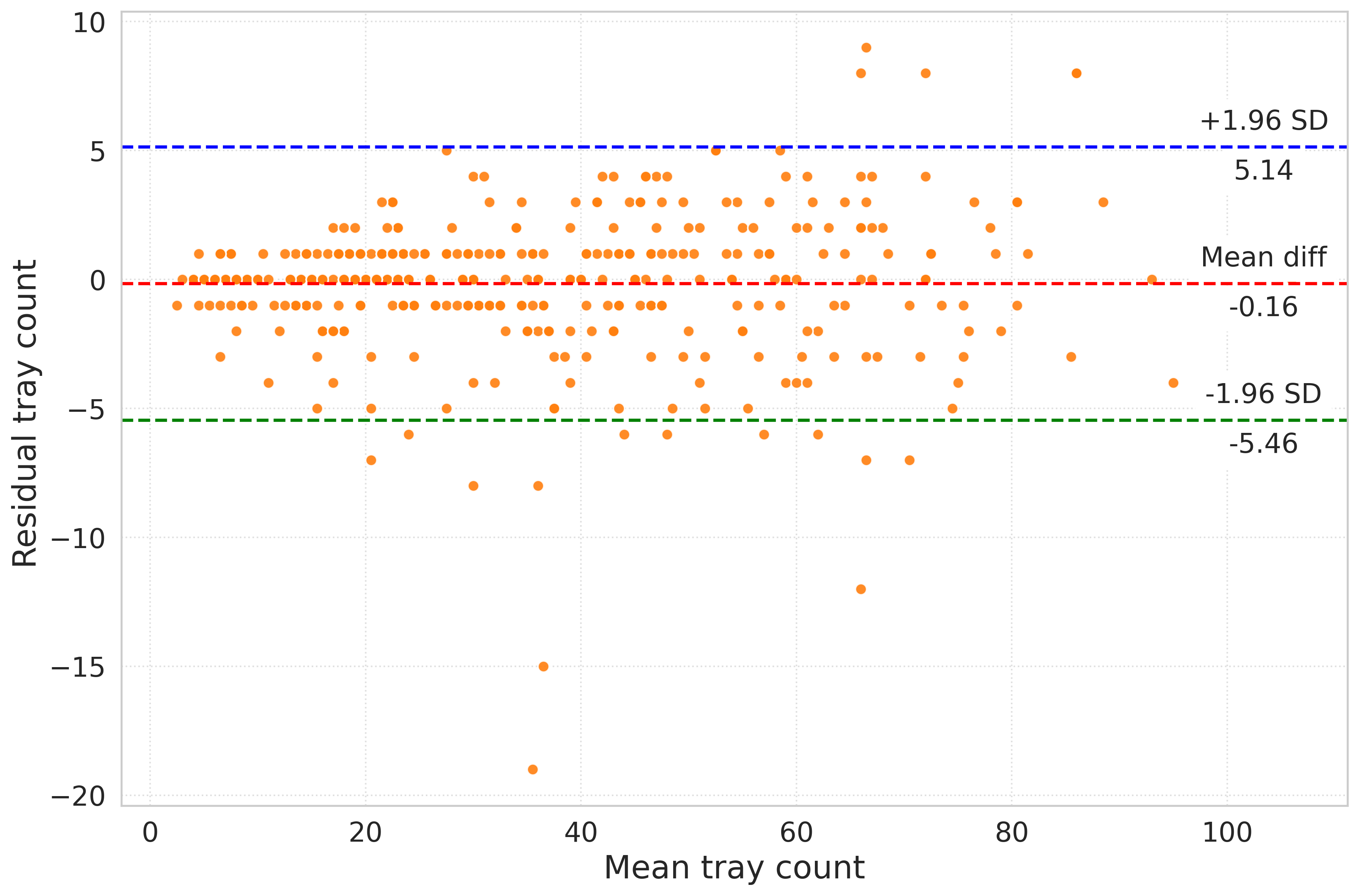}
        \includegraphics[width=0.48\textwidth]{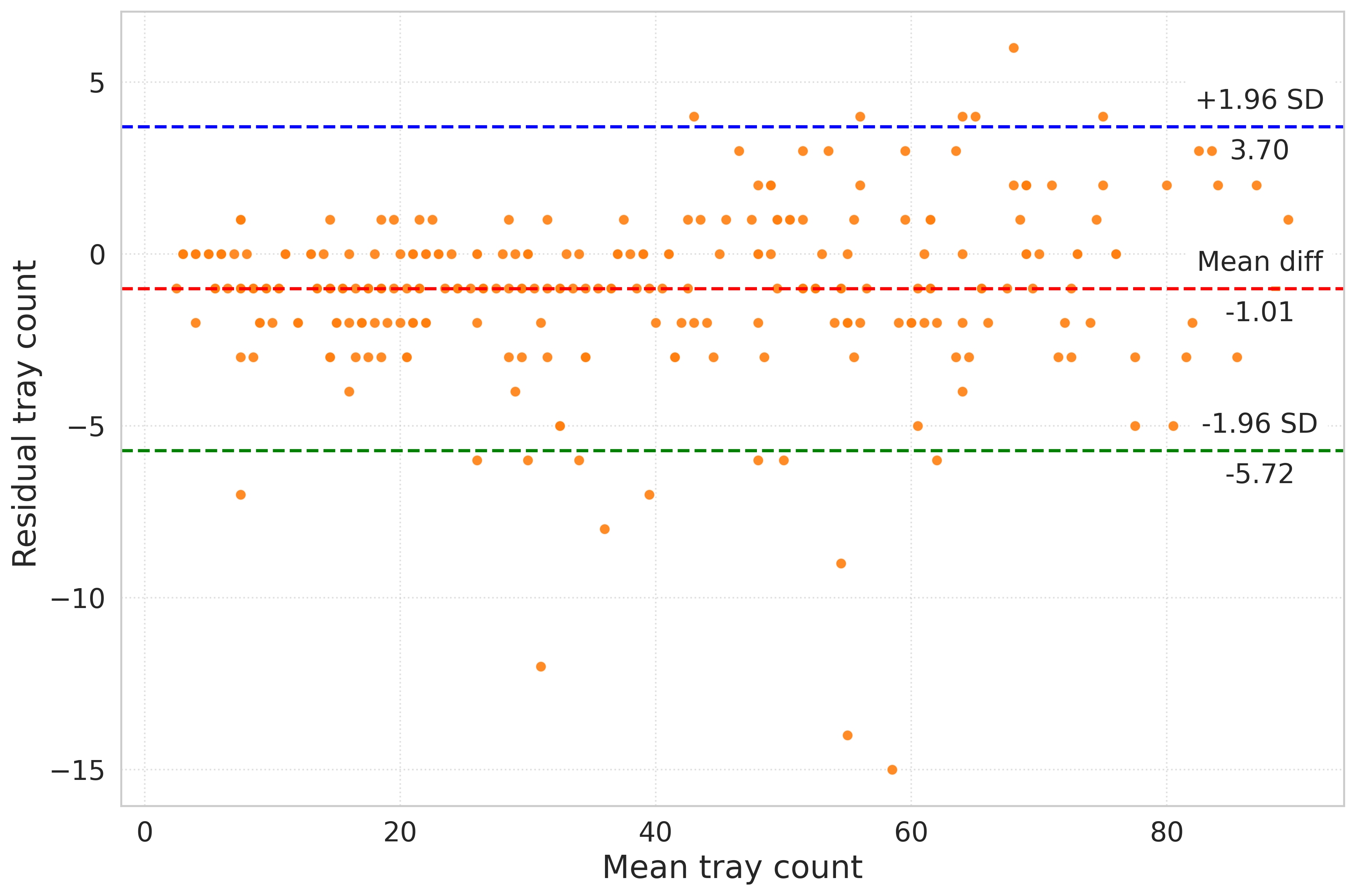}
    \caption{Bland-Altman plots comparing estimated and actual tray counts in Santa Maria and Salinas fields. (Left) Santa Maria field shows a mean underestimation bias of 0.16 trays with 95\% limits of agreement at $\pm$5.46 trays. (Right) Salinas field showing a mean underestimation bias of 1.01 trays with 95\% limits of agreement at $\pm$5.72 trays}
    \label{fig:bland_altman_plots_comparison}
\end{figure}

Figures \ref{fig:scatter_plots_comparison} presents scatter plots illustrating the correlation between the estimated and actual number of harvested trays per cart for the Santa Maria and Salinas fields. The red dashed line represents the ideal fit, indicating where estimated values perfectly align with actual values. A strong linear relationship was observed in both strawberry fields, with a high correlation (Pearson r = 0.99) between actual and estimated harvested trays. In both cases, most harvest tray data clustered closely around the ideal fit line. Nonetheless, the tray counts from several carts were underestimated or overestimated, indicating limitations in the data collection and processing phases.

Bland-Altman analysis was conducted in both fields to evaluate further the estimated yield based on tray count and the potential presence of system bias, as shown in Figure \ref{fig:bland_altman_plots_comparison} (Left). The Bland-Altman plot for Santa Maria showed a mean difference of -0.16 trays, indicating that the yield estimation pipeline, on average, underestimated the count by 0.16 trays when compared to the ground truth data. Similarly, the Bland-Altman plot for Salinas (Figure \ref{fig:bland_altman_plots_comparison} (Right)) showed a mean difference of -1.01 trays, also suggesting an average underestimation of 1.01 trays. Furthermore, for 95\% of the harvest data, the estimated tray count was likely to fall within 5.46 trays for Santa Maria and 5.72 for Salinas from the actual tray count.

\subsubsection{Yield map generation}
\label{sec:yield map generation}
\begin{figure}
    \centering
    \includegraphics[width=0.89\linewidth]{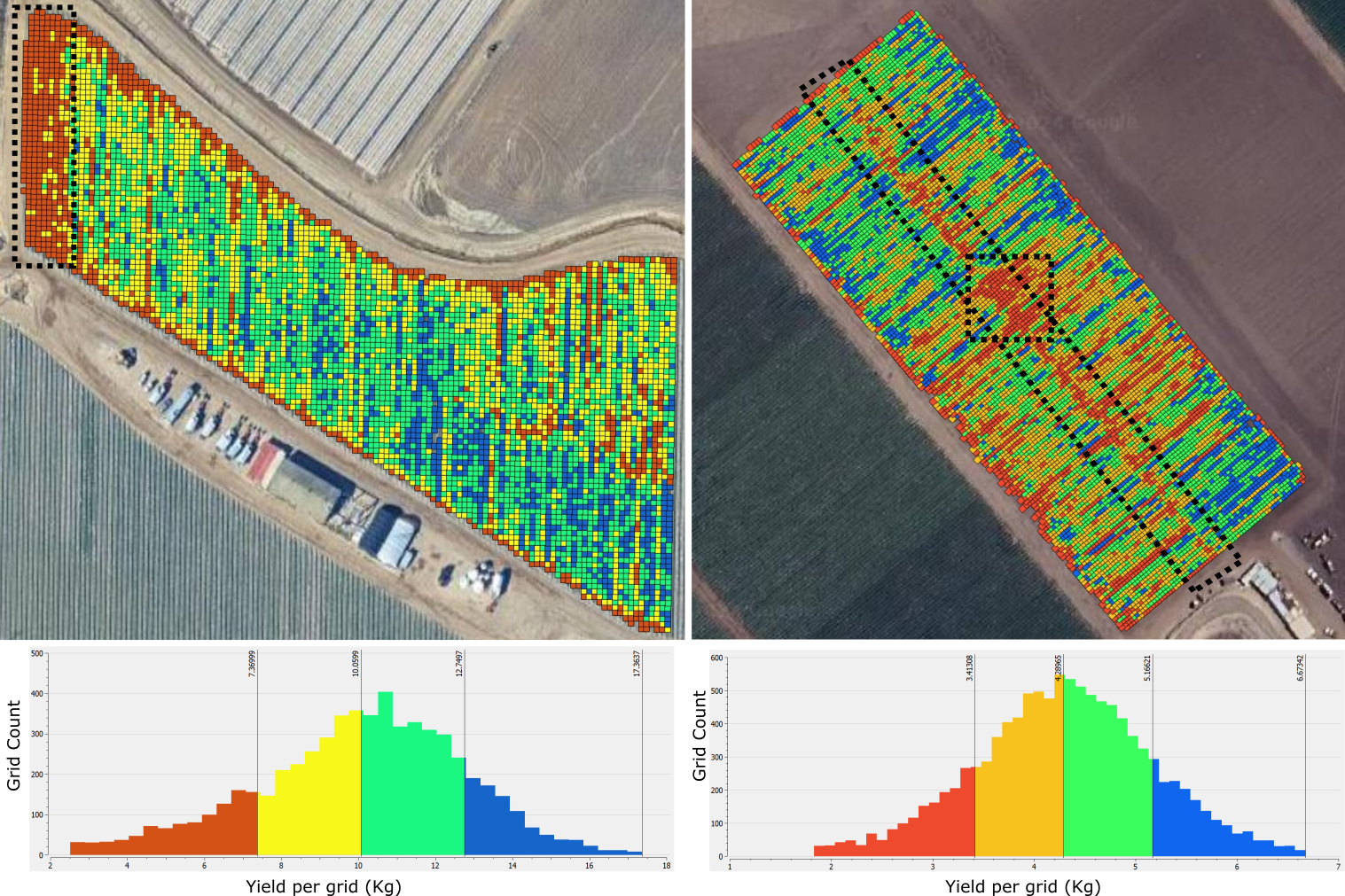}
    \caption{Generated cumulative yield maps of all harvest days in Santa Maria (Left) and season-long harvest in Salinas (Right) strawberry field. The histograms show the distribution of yield measured in kilograms per grid. Color coding categorizes the yield where red shows very low yield (\(\text{yield per grid} \leq \mu - \sigma\)), yellow shows below-average yield (\(\mu - \sigma < \text{yield per grid} \leq \mu\)), green shows above-average yield (\(\mu < \text{yield per grid} \leq \mu + \sigma\)), and blue shows very high yield (\(\text{yield per grid} > \mu + \sigma\)), where \(\mu\) is mean and \(\sigma\) is the standard deviation.}
    \label{fig:final_yield_map}
\end{figure}

Figure \ref{fig:final_yield_map} shows the generated yield map for the entire harvest period in Santa Maria and Salinas with a grid resolution equal to the row spacing (Santa Maria: 163 cm and Salinas: 122 cm). The color code represents the variability in yield intensity, defined by the mean and standard deviation. Yields within one standard deviation below the mean are indicated in yellow, while those above the mean are shown in green. Yields more than one standard deviation below the mean are marked in red, and those above are colored blue. The average yield per grid was 10.059 Kg (SD: 2.689) for Santa Maria and  4.289 Kg (SD: 0.877) for Salinas. This estimation corresponds to a yield of 3.785  $\text{Kg/m}^2$ (0.774 $\text{lbs/ft}^2$) for Santa Maria and 2.881  $\text{Kg/m}^2$ (0.590 $\text{lbs/ft}^2$) for Salinas respectively. For Santa Maria, it was observed that the yield in the left section of the field (marked by a black dotted box) was substantially lower than in other areas of the field. This drop in yield was related to a field section where plants died due to a soil-borne disease later in the season. 

Additionally, in Salinas, the yield at the center of the field (also marked by a black square box) was notably low. The seedlings in this area were found dead within a couple of weeks after planting and were not replanted. The low-yield sections in Salinas, indicated by a black rectangular dotted strip, align with the approximate center of the field from which pickers started harvesting. Additionally, low yield was observed in the field edges in both harvest locations. The grids were generated using a few key field coordinates (e.g., corners), and the grid cells at field edges may have partially overlapped with the actual cropped area. This issue is also present at the edges of the schematic representation of the yield map generation step in Figure \ref{fig:yield at desired resolution}. While these yield maps were analyzed for general trends observed in the field, they can be used for a more comprehensive understanding of spatial variability. Additionally, since the seasonal yield map was generated by accumulating yields from each harvest day, these maps can provide insights into temporal yield variations.

\subsection{Discussion}
The developed iCarritos and data processing pipeline showed promising potential for precision yield estimation and mapping in strawberry fields with high accuracy and reliability across different field conditions and strawberry varieties. The yield distribution estimation at the row segment level achieved a mean accuracy of 90.48\%. While the estimation had a slight underestimation bias of 13 grams, it was relatively small compared to the average tray weight of 4.25 Kg. Furthermore, evaluating the yield distribution at the tray level consistently estimated the full tray weight close to the ground truth with an average accuracy of 94.05\%. The yield distribution information is crucial for estimating the season-long yield and generating the yield map. Additionally, the system showed promising potential for yield estimation in large-scale multi-day harvesting scenarios. For Santa Maria, the estimated yield based on tray count per cart was 94.89\%, with an underestimation bias of 0.16 trays over 14 harvest sessions. Similarly, for Salinas, the estimated tray count accuracy was also high at 94.20\%, with an underestimation bias of -1.01 trays across 20 harvest sessions. The strong correlation (Pearson r = 0.99) between estimated and actual tray counts in both fields indicates that the system reliably estimated yield across different harvest days and locations. Furthermore, the generated yield maps would be valuable for precise, targeted field management throughout the season. 

The achieved level of accuracy in estimating yield and yield distribution is significant, given the various factors and challenges associated with GPS positional error and the manual nature of harvesting. We believe several factors contributed to such performance. Integrating the CNN-LSTM model for filtering non-picking data helped remove a substantial amount of irrelevant non-picking data during the harvest. The algorithms developed to address row completion and occupancy violations effectively identified and resolved the violations, mapping the relevant data to correct harvest locations. Without these algorithms, valuable data from the harvested row could have been lost. In contrast, neighboring rows might have shown substantially higher yields due to mistakenly assigned data from multiple carritos to a single row. Additionally, the use of adaptive polynomial regression for interpolating yield data effectively addressed the non-uniform data sampling due to the stop-and-harvest behavior of the pickers. 

The developed yield map can be used for early, targeted, and efficient field management, including addressing early disease outbreak detection and identifying irrigation system failures. The yield maps highlight high and low-yield areas within the fields. For example, the low-yield regions affected by disease outbreaks or plant death occurred in Santa Maria and Salinas. Additionally, analyzing the yield map for individual harvest days showed that some grids within the picking area had zero yields on particular harvest days. In Santa Maria, on average, 13.1\% of the grids had zero yield, while in Salinas, on average, 22.7\% of the grids had zero yield data. This could be due to various factors. Specifically in Salinas, the yield was significantly low due to a disease outbreak, and pickers had to switch between multiple rows (sometimes up to five) to fill a single tray. Since the pickers traveled longer distances to gather enough berries, some areas had zero yield. In both fields, the pickers may have missed some rows during harvesting, or there might not be any berries to harvest. Furthermore, the data was only recorded if the SBAS correction was available. If the GPS signal was lost in the middle of the harvest, pickers were not stopped, and the data during that period was lost. The pickers sometimes dropped the carritos from a height while moving and setting them up for harvesting. This led to loose connections, broken parts, power loss, or disconnections that resulted in occasional loss of harvest data. Furthermore, because of the extended harvesting periods (up to eight hours) in both fields, sometimes a carrito battery was discharged entirely in the middle of the harvest, and data was not recorded afterward. 

Despite these challenges, the proposed yield data collection, estimation, and mapping system has promising potential for low-cost, practical yield estimation and mapping. The flexibility to generate yield maps at various grid resolutions will be crucial for understanding yield variability in small and large areas due to microclimate, soil conditions, and plant health. Furthermore, the developed system has a low barrier to entry for existing harvesting operations. As mentioned earlier, the iCarritos were developed by modifying off-the-shelf picking carts. The pickers were not trained before using the iCarritos and were deployed on a commercial field with minimal disruption to the established harvesting practices.

\subsection{Limitations and Future Work}
Future work could focus on developing robust algorithms to reduce GPS localization errors by employing techniques such as predictive modeling using motion and velocity data and sensor fusion by combining data from multiple sensors (GPS, IMU, and potentially additional sensors). The yield estimation and mapping software could be improved with more robust algorithms to handle data loss, row completion, occupancy condition violations, or sensor malfunction. While the current study focused on system development and yield map generation, future research could expand into spatiotemporal yield variability analysis. This could include investigating the effects of different fumigation methods, treatment approaches, irrigation strategies, and crop rotation patterns. Furthermore, integrating the carrito yield data with other sensing modules, such as soil moisture sensors, weather stations, and crop imaging systems, could provide a more comprehensive understanding of the field and crops for informed data-driven decision-making.

\section{Conclusions}
\label{sec:conclusion}
In this study, instrumented picking carts were developed to collect real-time harvest data during commercial strawberry harvesting across two strawberry-growing regions in California. Algorithms were developed to estimate the yield and generate the yield map by addressing issues such as irrelevant non-picking data, harvest condition violation, and variability due to the manual nature of the harvesting. Based on the study, the following conclusions were drawn. 
\begin{itemize}
    \item Low-cost sensors, machine learning models, and effective noise-filtering and data interpolation techniques can accurately create high-resolution yield distribution information. The system achieved promising yield distribution estimation accuracy of 90.48\% based on row segment level evaluation and 94.05\% based on the tray level evaluation.
    
    \item The developed picking carts and data processing algorithms can be utilized for scalable and adaptable yield estimation and yield map generation under various field conditions, strawberry varieties, and planting architectures. Despite some minor underestimation biases, the system reliably estimated season-long yields based on tray count with over 94\% accuracy. The achieved accuracy and strong correlation (Pearson r=0.99) between estimated and actual tray counts demonstrate the system's reliability for season-long yield monitoring.
    
\end{itemize}

Future studies could focus on advancing sensing technologies and algorithms to reduce sensor localization errors and conducting spatiotemporal yield analysis in different scenarios to improve the accuracy of the estimated yield. Additionally, extending the application of the developed system to harvest other specialty crops with similar harvesting patterns could broaden the impact of the study.

\subsubsection*{Acknowledgments}
This work was funded by the USDA Agricultural Research Service (USDA-ARS) Areawide Pest Management Grant Program project ``Site-Specific Soil Pest Management in Strawberry and Vegetable Cropping Systems Using Crop Rotation and a Needs-Based Variable Rate Fumigation Strategy” through Non-Assistance Cooperative Agreements 58-2038-9-016 and 58-2038-3-029. We sincerely appreciate the invaluable contributions of Dennis Lee Sadowski, Alejandro Torres Orozco, and Clarence Codod, whose expertise and support were instrumental to the success of this study.

\bibliographystyle{apalike}

\end{document}